\documentclass{article}

\usepackage{amsmath}

\usepackage{caption}

\usepackage{microtype}
\usepackage{graphicx}
\usepackage{subfigure}
\usepackage{booktabs} 
\usepackage[english]{babel}

\usepackage{amsmath}

\usepackage{enumitem}

\usepackage{graphicx}
\usepackage{array}
\usepackage{adjustbox}

\usepackage{booktabs}
\usepackage{xcolor}
\usepackage{hyperref}

\usepackage[accepted]{icml2025}

\usepackage{amsmath}
\usepackage{amssymb}
\usepackage{mathtools}
\usepackage{amsthm}

\theoremstyle{plain}

\theoremstyle{definition}

\theoremstyle{remark}

\usepackage[textsize=tiny]{todonotes}

\icmltitlerunning{ReMiDi: Reconstruction of Microstructure Using a Differentiable Diffusion MRI Simulator}

\begin{document}

\twocolumn[
\icmltitle{ReMiDi: Reconstruction of Microstructure Using a Differentiable Diffusion MRI Simulator}



\icmlsetsymbol{equal}{*}

\begin{icmlauthorlist}
\icmlauthor{Prathamesh Pradeep Khole}{ucsc}
\icmlauthor{Zahra Kais Petiwala}{ucsc}
\icmlauthor{Shri Prathaa Magesh}{iitm}
\icmlauthor{Ehsan Mirafzali}{ucsc}
\icmlauthor{Utkarsh Gupta}{ucsc}
\icmlauthor{Jing-Rebecca Li}{inria}
\icmlauthor{Andrada Ianus}{kcl}
\icmlauthor{Razvan Marinescu}{ucsc}
\end{icmlauthorlist}

\icmlaffiliation{ucsc}{Department of Computer Science and Engineering, University of California Santa Cruz, Santa Cruz, USA}
\icmlaffiliation{iitm}{Indian Institute of Technology Madras, Chennai, India}
\icmlaffiliation{kcl}{School of Biomedical Engineering and Imaging Sciences, King's College London, London, UK}
\icmlaffiliation{inria}{Equipe Idefix, Inria-Saclay, Palaiseau, France}

\icmlcorrespondingauthor{Prathamesh Pradeep Khole}{pkhole@ucsc.edu}

\icmlkeywords{Medical Imaging, Differentiable Simulator, Auto-Encoder, Deep Learning, Latent Representation, Reconstruction}

\vskip 0.3in
]



\printAffiliationsAndNotice  
\newcommand{\argmax}{\operatorname{argmax}}
\newcommand{\argmin}{\operatorname{argmin}}

\begin{abstract}
We propose ReMiDi, a novel method for inferring neuronal microstructure as arbitrary 3D meshes using a differentiable diffusion Magnetic Resonance Imaging (dMRI) simulator. We first implemented in PyTorch a differentiable dMRI simulator that simulates the forward diffusion process using a finite-element method on an input 3D microstructure mesh. To achieve significantly faster simulations, we solve the differential equation semi-analytically using a matrix formalism approach. Given a reference dMRI signal $S_{ref}$, we use the differentiable simulator to iteratively update the input mesh such that it matches $S_{ref}$ using gradient-based learning. Since directly optimizing the 3D coordinates of the vertices is challenging, particularly due to ill-posedness of the inverse problem, we instead optimize a lower-dimensional latent space representation of the mesh. The mesh is first encoded into spectral coefficients, which are further encoded into a latent $\textbf{z}$ using an auto-encoder, and are then decoded back into the true mesh. We present an end-to-end differentiable pipeline that simulates signals that can be tuned to match a reference signal by iteratively updating the latent representation $\textbf{z}$. We demonstrate the ability to reconstruct microstructures of arbitrary shapes represented by finite-element meshes, with a focus on axonal geometries found in the brain white matter, including bending, fanning and beading fibers. Our source code is available \href{https://github.com/BioMedAI-UCSC/ReMiDi}{online}.

\end{abstract}

\section{Introduction}
\label{submission}

Diffusion Magnetic Resonance Imaging (dMRI) \cite{LeBihan1986} has revolutionized our ability to non-invasively probe tissue microstructure, offering crucial insights for diagnosing and monitoring neurological disorders such as stroke, multiple sclerosis and cancers \cite{bodini2009exploring}. The current paradigm of estimating tissue properties from the dMRI signals involves fitting parametric models such as 3D diffusion tensors \cite{Basser1994}, or simple multi-compartment models which include the signal contribution from different tissue environments (e.g. NODDI\cite{zhang2012noddi}, WMTI \cite{BENITEZ201464}, SANDI \cite{PALOMBO2020116835}, etc.). Such models are typically limited to simple diffusion patterns, such as unidimentional diffusion inside the axons, thus cannot represent complex tissue microstructures. Estimating more complex tissue features, such as axonal undulation, bending or beading has been elusive, in part due to the lack of a flexible framework suitable for solving this inverse problem.  


Diffusion MRI simulators are a crucial tool for calculating the dMRI signal corresponding to various underlying tissue microstructures, usually represented by 3D meshes \cite{Palombo2022, CALLAGHAN2020117107, Ianus2021}. o frameworks such as Disimpy \cite{Kerkelä2020} and Camino \cite{Cook2006Camino} simulate the diffusion signal by integrating random walks of many particles, yet the stochasticity and the discontinuities in particle velocities near the boundary conditions make it difficult to define differentiability in the classical sense and efficiently compute it. On the other hand, the latter class of simulators, such as SpinDoctor \cite{Li2019SpinDoctor}, solve the Bloch-Torrey Partial Differential Equation (BTPDE) \cite{PhysRev.104.563} on pre-defined finite-element meshes \cite{HOLE198827}. Given the smoothness of the BTPDE solutions, differentiability is possible, yet we are not aware of any differentiable implementations of SpinDoctor or similar BTPDE-based simulators. 

To enable inference of complex microstructure using such powerful simulators, we reimplemented SpinDoctor into a fully differentiable dMRI simulator using PyTorch \cite{Paszke2019}, enabling end-to-end gradient computation. In addition, to enable a fast computation of the forward simulation, we implemented the matrix formalism \cite{CALLAGHAN199774} method for solving the BTPDE, which approximates the mesh into a semi-analytical model which is solved by the eigendecomposition of a large Finite Element Method (FEM) \cite{Turner1956} matrix. Our key contributions include:

\begin{enumerate}
    \item A differentiable implementation of SpinDoctor \cite{Li2019SpinDoctor}, a BTPDE-based dMRI simulator that allows for gradient-based optimization techniques and seamless integration with deep neural networks.
    \item A fast implementation of the matrix formalism, including the integration of $\xi$-torch \cite{Kasim2020xiTorch} to solve the generalized eigenvalue problem in an efficient and differentiable way.
    \item A novel technique for solving inverse problems in dMRI using stochastic gradient descent by backpropagating through the BTPDE simulator.
    \item The use of a spectral auto-encoder \cite{LEMEUNIER2022131} to optimize a lower-dimensional representation of the input mesh represented by spectral coefficients. This approach reduced the degrees of freedom and alleviates the ill-posedness of the problem, further removing the tight coupling between vertices and enabling the optimisation of orthogonal spectral coefficients.
    \item Comprehensive experiments with different meshes as well as ablation studies, and comparison with a neural network as a baseline, highlighting the ability of our model to capture bends and beading of the input meshes. 
\end{enumerate}

\begin{figure*}[ht!]
\begin{center}
\vskip 0.2in
\centerline{\includegraphics[width=0.9\textwidth]{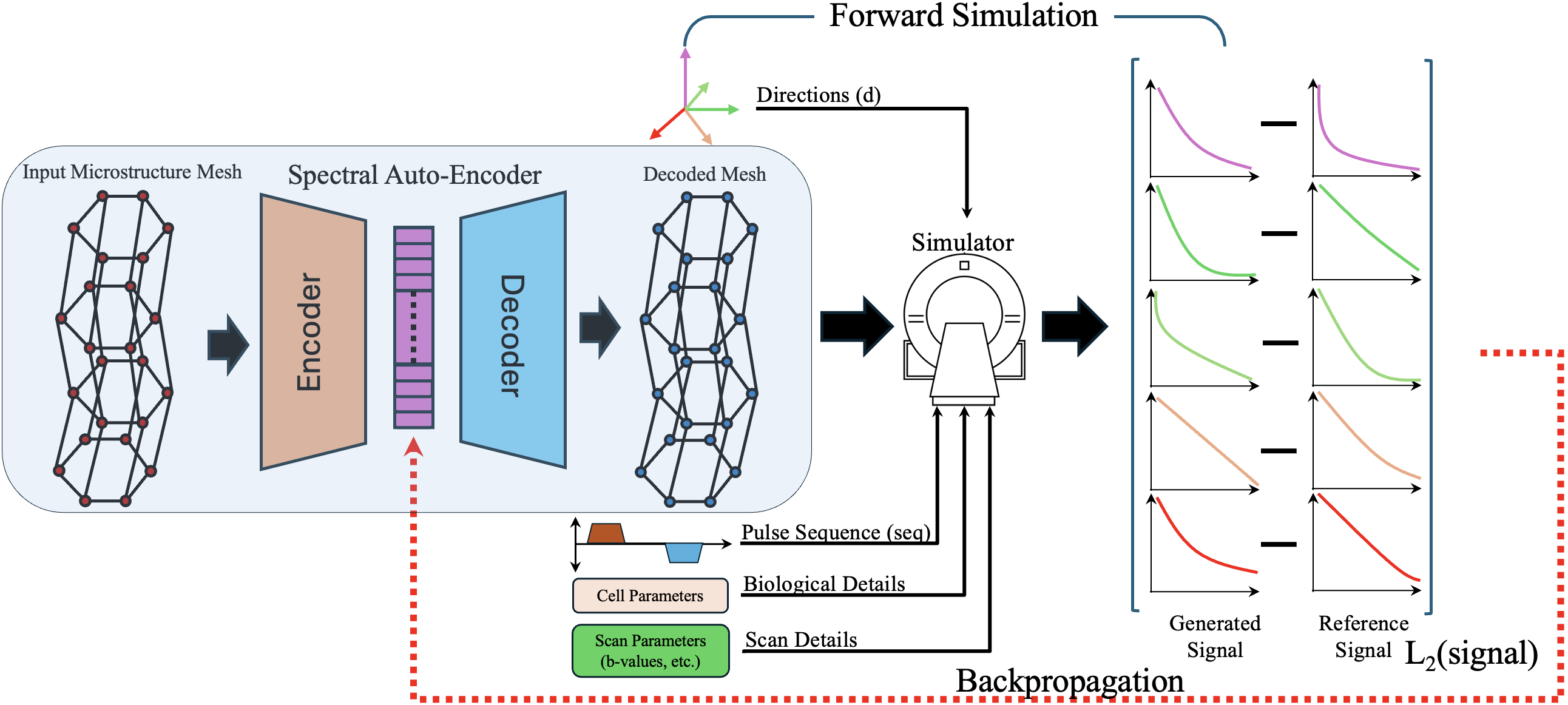}}
\caption{Overview of the ReMiDi-based dMRI reconstruction pipeline. A 3D mesh decoded by a Spectral Auto-Encoder (SAE) is given as input to the differentiable dMRI simulator. The final loss function between the simulated dMRI signal and a reference signal is backpropagated to the latent of the SAE, and the mesh is iteratively updated with gradient-based learning.}
\label{ReMiDi_pipeline}
\end{center}
\vskip -0.2in
\end{figure*}

\section{Background}

\subsection{Bloch-Torrey Partial Differential Equation}

\par{The Bloch-Torrey Partial Differential Equation (BTPDE) provides a mathematical framework for modeling the evolution of complex transverse water proton magnetization in the rotating frame under a time-varying magnetic field gradient \cite{PhysRev.104.563, Li2019SpinDoctor}. It models the evolution of the magnetization $M_i(\vec{x}, t)$ at position $\vec{x}$ and time $t$ in each compartment $\Omega_i$ within a domain $\Omega$ as\footnote{$\Omega$ generally represents the entire voxel being simulated, while a compartment $\Omega_i$ can represent a neuron, an axon, or the nucleus of a neuron}:}
\[
\frac{\partial M_i}{\partial t} = -\left(-\nabla \cdot D_i \nabla + \frac{1}{T_{2,i}} + \hat{\i} \gamma \vec{g}(t) \cdot \vec{x} \right) M_i,
\]
where $M_i = M_i(\vec{x}, t)$ represents the transverse magnetization, $\vec{x} \in \Omega_i$ and $-\nabla \cdot D_i \nabla$ is the generalized Laplace operator with $D_i$ as the diffusion tensor that characterizes the movement of water molecules in $\Omega_i$. $T_{2,i}$ denotes the transverse relaxation time $T_2$ for compartment $\Omega_i$, which accounts for the decay of transverse magnetization due to spin-spin interactions. $\hat{\i}$ is the imaginary unit and $\gamma = 2.675 \times 10^8$ rad s$^{-1}$ T$^{-1}$ is the gyromagnetic ratio for water protons. The time-varying magnetic field gradient $\vec{g} : [0, T_{\text{echo}}] \to \mathbb{R}^3$ encodes spatial information by introducing a position-dependent variation in the precession frequency. In Pulsed Gradient Spin Echo (PGSE) sequences, the gradient is applied as two short pulses separated by a diffusion time $\Delta$, allowing the measurement of water diffusion within a specific time scale.


Boundary conditions govern the magnetization flux across compartment interfaces ($\Gamma_{ij}$) and the domain's outer boundary ($\partial \Omega$), determined by permeability coefficients ($\kappa_{ij}$), which regulate the exchange of magnetization, and equilibrium factors ($c_{ij}$), which establish the spin density balance between compartments.

The initial condition is given by $M_i(\vec{x}, 0) = \rho_i,$ where $\rho_i$ is the initial spin density, representing the number of protons contributing to the signal in $\Omega_i$. The diffusion MRI signal at the echo time $T_{\text{echo}}$ is computed as:
\[
S(\vec{g}) = \int_\Omega M(\vec{x}, T_{\text{echo}}) \, d\Omega(\vec{x}),
\]
where $S(\vec{g})$ represents the measured MRI signal.
The BTPDE accounts for non-Gaussian diffusion effects, microstructural anisotropy, and complex tissue interactions, providing a detailed representation of the physical processes underlying diffusion MRI.

\subsection{Matrix Formalism Solution for Bloch-Torrey PDE}

Matrix formalism \cite{CALLAGHAN199774, Li2019SpinDoctor} provides a powerful approach to approximate solutions for BTPDE by projecting it onto a truncated basis of eigenfunctions of the Laplace operator. The generalized Laplace operator $-\nabla \cdot D \nabla$, defined over the domain $\Omega = \bigcup_{i=1}^{N_{\text{cmpt}}} \Omega_i$, is characterized by its eigenvalues and eigenfunctions. Let $\{(\phi_n, \lambda_n)\}_{n \in \mathbb{N}^*}$ denote the $L^2$-normalized eigenfunctions $\phi_n$ and the corresponding eigenvalues $\lambda_n$ of the operator. For each compartment $\Omega_i$, the eigenfunctions satisfy:
\[
-\nabla \cdot D_i \nabla \phi_i(\vec{x}) = \lambda \phi_i(\vec{x}), \quad \vec{x} \in \Omega_i,
\]
with boundary conditions consistent with the BTPDE, ensuring flux continuity and permeability at interfaces. 

Using the eigenfunctions $N_{\text{eig}}$ with the smallest eigenvalues, the magnetization $M(\vec{x}, t)$ can be projected onto this truncated basis as:
\[
M^{MF}(\vec{x}, t) \approx \vec{\phi}^T(\vec{x}) \vec{\nu}(t),
\]
where $\phi(\vec{x}) = [\phi_1(\vec{x}), \dots, \phi_{N_{\text{eig}}}(\vec{x})]^T$ is the vector of eigenfunctions, and $\nu(t) = [\nu_1(t), \dots, \nu_{N_{\text{eig}}}(t)]^T$ are the time-dependent coefficients. Substituting this expansion into the BTPDE leads to a system of ordinary differential equations for $\nu(t)$:
\[
\frac{d \vec{\nu}}{dt} = -K(\vec{g}(t)) \vec{\nu}(t),
\]
where $K(g(t)) = L + T + i\gamma A(g(t))$ is the complex-valued Bloch-Torrey operator in the eigenfunction basis. Here:
\begin{itemize}
    \item $L = \text{diag}(\lambda_1, \lambda_2, \dots, \lambda_{N_{\text{eig}}})$ is the diagonal matrix of Laplace eigenvalues,
    \item $T_{mn} = \int_\Omega \frac{1}{T_2(\vec{x})} \phi_m(\vec{x}) \phi_n(\vec{x}) \, d\Omega(\vec{x})$ is the transverse relaxation matrix, and
    \item $A(g(t)) = g_x A_x + g_y A_y + g_z A_z$, where $A_x$, $A_y$, and $A_z$ are matrices encoding the spatial gradients of the eigenfunctions.
\end{itemize}
The diffusion MRI signal for a given gradient $\vec{g}$ and a time-dependent function $f(t)$, representing the temporal profile of the gradient, is computed as:
\[
S^{\text{MF}}(f, \vec{g}) = \vec{\nu}^T(T_{\text{echo}}) \int_\Omega \vec{\phi}(\vec{x}) \, d\Omega(\vec{x}).
\]
In this context, $f(t) \in \{-1, 0, 1\}$ defines the temporal modulation of the gradient during the pulse sequence, such as in PGSE sequences. For these sequences, the gradient is expressed as $\vec{g}(t) = f(t) \vec{g}_d$, where $\vec{g}_d$ represents the fixed gradient direction and magnitude.

This formulation leverages matrix formalism by compressing the BTPDE into a reduced eigenfunction space, where the coefficients $\vec{\nu}(T_{\text{echo}})$ are computed. By reducing the problem to the most significant eigenfunctions of the Laplace operator, matrix formalism significantly simplifies the computation and leads to faster simulation times, necessary for our gradient-based learning approach.

\subsection{Discretization of Solution over Finite Element Mesh}

The Finite Element Method \cite{Turner1956} provides a systematic way to approximate the solution to the BTPDE by discretizing the domain \( \Omega \) into a finite set of elements. The continuous problem is projected onto a finite-dimensional subspace spanned by nodal basis functions \( \{\varphi_k\}_{k=1}^{N_{\text{node}}} \), where \( \varphi_k : \Omega \to [0, 1] \) are piecewise linear functions associated with the finite element nodes \( \{q_k\}_{k=1}^{N_{\text{node}}} \). The solution to the BTPDE is expressed as a linear combination of the basis functions:
\[
M(\vec{x}, t) = \sum_{k=1}^{N_{\text{node}}} \xi_k(t) \varphi_k(\vec{x}) = \xi^T(t) \varphi(\vec{x}),
\]
where \( \xi(t) = [\xi_1(t), \dots, \xi_{N_{\text{node}}}(t)]^T \) is the vector of time-dependent coefficients, and \( \varphi(\vec{x}) = [\varphi_1(\vec{x}), \dots, \varphi_{N_{\text{node}}}(\vec{x})]^T \) is the vector of basis functions. Substituting this representation into the BTPDE and applying the Galerkin method \cite{Johnson2009, Reddy2006} results in the following system of ordinary differential equations (ODEs): 
\[
M \frac{d\xi}{dt} = -\left(S + Q + R + i\gamma J(\vec{g}(t))\right) \xi(t),
\]
where the matrices \( M, S, Q, R, \) and \( J(\vec{g}(t)) \) encode specific physical properties of the system. The mass matrix \( M \) captures the overlap between finite element basis functions and ensures that the discrete representation of the solution conserves mass. The stiffness matrix \( S \) encodes the diffusion properties, reflecting how gradients of the basis functions interact through the diffusion tensor \( D(\vec{x}) \). The flux matrix \( Q \) incorporates the effects of permeable interfaces and boundary conditions, enabling proper treatment of compartmental interactions. The relaxation matrix \( R \) accounts for \( T_2 \)-relaxation effects, modulating the decay of transverse magnetization based on the relaxation properties of the tissue. Finally, the gradient matrix \( J(\vec{g}(t)) \) represents the effects of applied magnetic field gradients, with \( J_x, J_y, \) and \( J_z \) encoding spatial dependencies through coordinate-weighted mass matrices.

The FEM discretization converts the continuous operator equations into a set of matrix equations, which can be efficiently solved numerically. The resulting system describes the time evolution of the coefficients \( \xi(t) \), capturing the magnetization within the finite-dimensional subspace.

To compute the diffusion MRI signal, the solution \( \xi(t) \) at echo time \( T_{\text{echo}} \) is combined with the finite element basis functions. The signal is given by:
\[
S = \xi^T(T_{\text{echo}}) \int_\Omega \varphi(\vec{x}) \, d\Omega(\vec{x}).
\]
The finite element discretization focuses on breaking down the continuous domain into manageable elements, enabling precise numerical solutions for the BTPDE while ensuring computational efficiency and accurate handling of discretized physical properties.

\section{Methods}

\subsection{Differentiable Simulator}

\begin{figure}[ht]
\vskip 0.2in
\begin{center}
\centerline{\includegraphics[width=0.9\columnwidth]{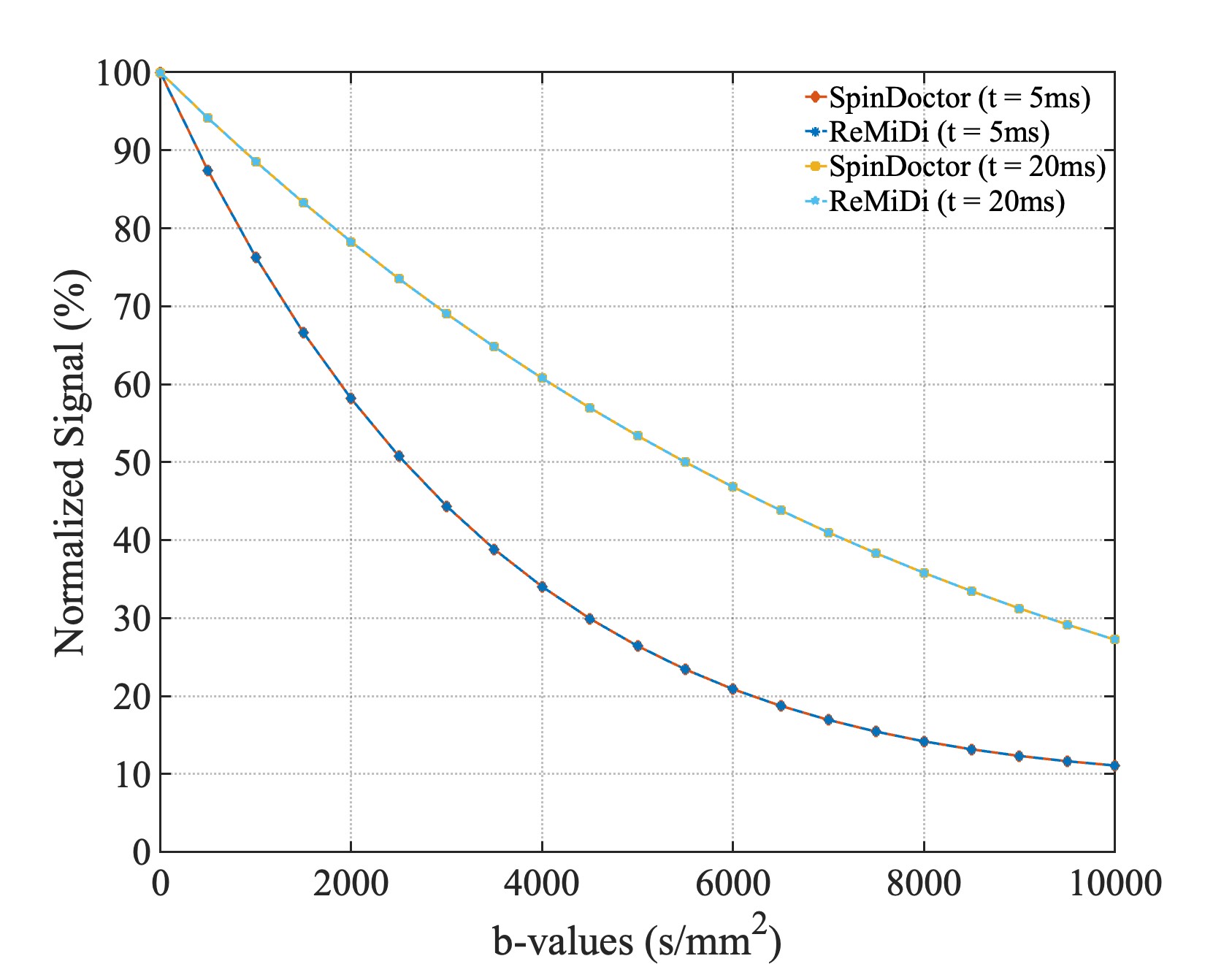}}
\caption{The signal generated by ReMiDi for a bent axon closely matches SpinDoctor's MATLAB implementation. As opposed to SpinDoctor, ReMiDi is implemented in PyTorch and is fully differentiable.}
\label{SpinDoctor_vs_ReMiDi}
\end{center}
\vskip -0.2in
\end{figure}

Simulation of dMRI signal generation involves solving the BTPDE for a given gradient sequence and microstructural substrate defined by a FE mesh. SpinDoctor \cite{Li2019SpinDoctor}, an FEM-based solver, accurately models the physics of diffusion and magnetic flux evolution but is limited to forward simulations, generating dMRI signals from a given mesh. It is not designed to solve the inverse problem of reconstructing a microstructure from a dMRI signal and lacks the differentiable framework needed for solving this efficiently with gradient-based learning.

To address this, we introduce ReMiDi, a differentiable dMRI simulator based on SpinDoctor but implemented in PyTorch. By using PyTorch's auto-differentiation capabilities and GPU-accelerated libraries like Xi-Torch for efficient eigenvalue decomposition, ReMiDi allows backpropagation through the simulation process. This enables us to minimize the error between simulated and reference signals by updating the input FE mesh. The output signal generated by ReMiDi matches the values from the original SpinDoctor implementation as shown in figure~\ref{SpinDoctor_vs_ReMiDi} for single bent axon.

In our simulations, we backpropagate only to the mesh vertices, keeping all other scan parameters—such as b-values, sequences, and microstructural properties—fixed. The backward pass updates the mesh to minimize the error between the generated and reference signals, focusing the optimization of the mesh geometry.

Mathematically, for an input FE mesh $\mathcal{M} \in \mathbb{R}^{V \times 3}$, where $V$ is the number of vertices, the simulator computes the dMRI signal $\hat{\mathcal{S}}$ as:
\[
\hat{\mathcal{S}} = \mathcal{F}(\mathcal{M}),
\]
where $\mathcal{F}(\mathcal{M})$ represents the forward simulation. The final optimization problem becomes: 
\[
\mathcal{M}^* = \argmin_{\mathcal{M}} \|\mathcal{F}(\mathcal{M}) - \mathcal{S}_{\text{ref}}\|_2^2.
\]
which is solved with gradient descent on $\mathcal{M}$.

\subsection{Reducing Ill-Posedness by Projecting to Latent Space}


The inverse problem of reconstructing a mesh from a dMRI signal is ill-posed because multiple microstructures can produce the same signal. To address this, we instead optimize a lower-dimensional representation $\mathbf{z}$ of the 3D mesh obtained using a Spectral Auto-Encoder (SAE) \cite{LEMEUNIER2022131} trained on a dataset of plausible axon section meshes. Instead of backpropagating directly to the mesh vertices, ReMiDi backpropagates to the latent vector $\mathbf{z}$. This approach is motivated by the work of \cite{Zhao2022}, who demonstrated that solving PDE-constrained inverse problems benefits from learned priors in a latent space, as it provides regularization, improves optimization stability, and enhances computational efficiency. We found that a dimension of 16 for $\mathbf{z}$ ensures that the reconstructed meshes are both smooth and flexible, and improve the stability and speed of the optimization.

Mathematically, the forward simulation in latent space is:
\[
\hat{\mathcal{S}} = \mathcal{F}(\mathcal{D}(\mathbf{z})),
\]
where $\mathcal{D}(\mathbf{z})$ is the SAE decoder that maps the latent vector $\mathbf{z}$ back to a mesh $\mathcal{M}$. The optimization problem becomes:
\[
\mathbf{z}^* = \argmin_{\mathbf{z}} \|\mathcal{F}(\mathcal{D}(\mathbf{z})) - \mathcal{S}_{\text{ref}}\|_2^2.
\]
and the optimized mesh is obtained as $\mathcal{M}^* = \mathcal{D}(\mathbf{z})$. Using the SAE reduces the problem's dimensionality, limits the search space to plausible meshes, and enhances the reconstruction's stability and speed. The full mesh reconstruction with ReMiDi is illustrated in Figure~\ref{ReMiDi_pipeline} and outlined in Algorithm~\ref{alg:RemiDi}.




\begin{algorithm}[tb]
   \caption{ReMiDi Reconstruction Method}
   \label{alg:RemiDi}
\begin{algorithmic}
   \STATE {\bfseries Input:} Input mesh $\mathcal{M} \in \mathbb{R}^{315 \times 3}$, loss multiplier $k$, learning rate $\eta$, reference signal $\mathcal{S}_{\text{ref}}$
   \STATE Initialize latent vector $\mathbf{z} = f_E(\mathcal{M})$
   \WHILE{not converged}
   \STATE Reconstruct mesh: $\mathcal{M} = f_D(\mathbf{z})$
   \STATE Simulate signal: $\hat{\mathcal{S}} = \mathcal{F}(\mathcal{M})$
   \STATE Compute loss: $\mathcal{L} = k \times \|\hat{\mathcal{S}} - \mathcal{S}_{\text{ref}}\|_2^2$
   \STATE Backpropagate gradients: $\frac{\partial \mathcal{L}}{\partial \mathbf{z}} = \frac{\partial \mathcal{L}}{\partial \hat{\mathcal{S}}} \cdot \frac{\partial \mathcal{S}(f_D(\mathbf{z}))}{\partial \mathbf{z}}$
   \STATE Update latent vector: $\mathbf{z} \leftarrow \mathbf{z} - \eta \frac{\partial \mathcal{L}}{\partial \mathbf{z}}$
   \ENDWHILE
   \STATE {\bfseries Output:} Reconstructed mesh $\mathcal{M}^*$
\end{algorithmic}
\end{algorithm}


\section{Experiments}

To demonstrate the effectiveness of ReMiDi, we used meshes with $V = 315$ vertices, each represented by 3D coordinates, resulting in an input size of (315, 3). To preserve geometric structure, 300 spectral coefficients were used in the SAE, while the latent space was set to 16x1, achieving around 20X reduction in the mesh representation.

The SAE was trained on datasets tailored to three types of deformation, which reflect the structural changes axons undergo during growth or under external forces: bending and twisting, fanning, and beading. Bending refers to the elongation and curving of axons caused by external mechanical forces, including fluid flow or compression from surrounding structures. Fanning involves slight, divergence-like bending without significant twisting, often seen near the cortical regions. Beading represents localized enlargements along the axons, commonly linked to injuries, modeled as sinusoidal surface transformations. (See Appendix for detailed descriptions.)

The SAE encoder and decoder were designed to learn downsampling and upsampling kernels with feature dimensions [300, 64, 32, 16]. Training was conducted in batches of 16 for efficiency, while validation and testing utilized batch sizes of 64. To further enhance the dataset, we applied scaling factors of 0.5x and 1.5x to the meshes along the X, Y and Z axes, increasing the diversity of training data. Additionally, for all experiments, we performed additional rotations of the meshes to increase the number and diversity of samples: $90^{\circ}$ rotations around each axis individually, $90^{\circ}$ rotations around combinations of two axes, a $90^{\circ}$ rotation around all three axes simultaneously, and $180^{\circ}$ rotations around each axis individually. This augmentation ensured the robustness of the model and avoided data imbalance, as the meshes were scaled and rotated differently along various axes. Importantly, all experiments were performed using meshes that the auto-encoder had not previously encountered.


\begin{algorithm}[tb]
   \caption{Chamfer Distance Computation}
   \label{alg:chamfer}
\begin{algorithmic}
   \STATE {\bfseries Require:} $P_1 = \{x_i \in \mathbb{R}^3\}_{i=1}^n$, $P_2 = \{x_j \in \mathbb{R}^3\}_{j=1}^m$
   \STATE {\bfseries Compute:} Chamfer distance between $P_1$ and $P_2$
   \STATE Forward Distance $d_{1} \leftarrow \frac{1}{2n}\sum_{i=1}^n \vert x_{i} - \mathrm{NN}(x_{i}, P_{2})\vert$
   \STATE Backward Distance $d_{2} \leftarrow \frac{1}{2m}\sum_{j=1}^m \vert x_{j} - \mathrm{NN}(x_{j}, P_{1})\vert$
   \STATE {\bfseries return} $d_{1} + d_{2}$
\end{algorithmic}
\end{algorithm}

To evaluate the mesh reconstruction, we use the Chamfer distance \cite{Barrow1977}, defined between two point clouds $P_1$ and $P_2$ as shown in Algorithm~\ref{alg:chamfer}, where $\text{NN}(x,P)=\argmin_{x'\in P}\|x - x'\|$ is the nearest neighbor function. The Chamfer distance is preferred over RMSE as it does not require point-to-point correspondence and is more robust to local deformations, allowing better comparison of overall shape similarity even when vertex orderings differ or point clouds have different densities. Additionally, to account for the fact that the dMRI signal is invariant to reflections around the center point of the bounding box and the invariance of the dMRI signal to certain rotations when measuring symmetric meshes, we report a modified Chamfer distance $d'$ that accounts for both inversions and rotations, computed as $d' = \min_{R \in \mathcal{R}} (d(R(\mathcal{M}), \mathcal{M}_{\text{ref}}))$, where $d$ is the standard Chamfer distance, and $R \in \mathcal{R}$, where $\mathcal{R}$ represents all possible rotations around each axis and inversions around the center point. 




\section{Results}

\begin{figure*}[ht!]
\vskip 0.2in
\begin{center}
\centerline{\includegraphics[width=0.95\textwidth]{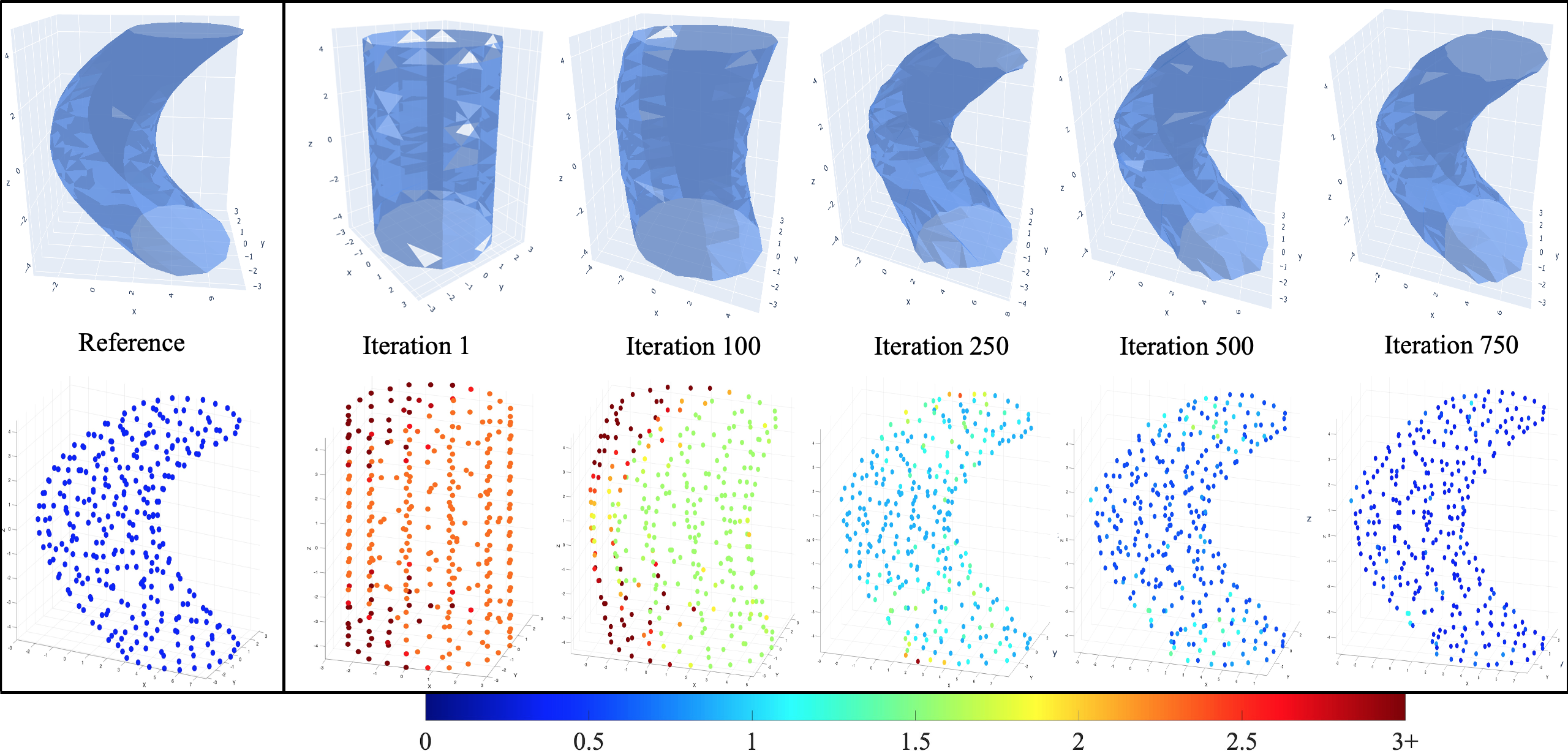}}
\caption{Overview of iterative reconstruction of a bent axon by ReMiDi at different gradient descent iterations. The ground-truth mesh is shown to the left. Top row: Evolution of the triangulated mesh surface. Bottom row: Corresponding point cloud representations of mesh vertices.  The color scale shows Chamfer distance (lower values are better).}
\label{Iterative_reconstruction_steps}
\end{center}
\vskip -0.2in
\end{figure*}

\begin{figure}[ht]
\vskip 0.2in
\begin{center}
\centerline{\includegraphics[width=0.9\columnwidth]{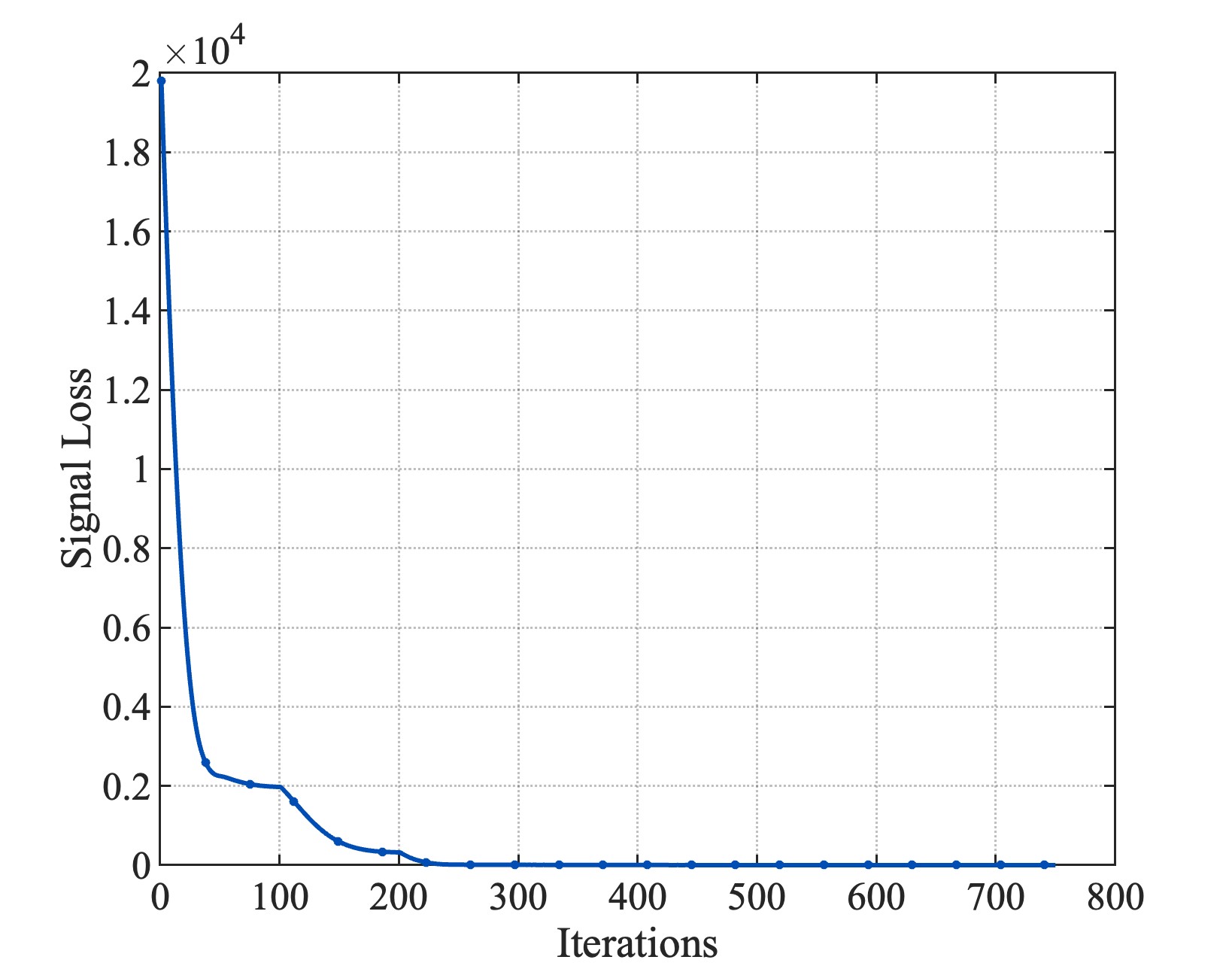}}
\caption{dMRI signal loss over iterations for reconstruction of a bent axon using ReMiDi}
\label{signal_loss_reconstruction}
\end{center}
\vskip -0.2in
\end{figure}

Fig.~\ref{Iterative_reconstruction_steps} and~\ref{signal_loss_reconstruction} illustrate ReMiDi's iterative reconstruction process. While starting with a straight cylinder, ReMiDi gradually refines the mesh based on the dMRI signal, and by iteration 250, the general curved shape is clearly observed. By the end of iteration 750, we notice that there are no local regions of the mesh with significantly high errors (bottom plots in Fig ~\ref{Iterative_reconstruction_steps} show that all vertices have low error), and the mesh broadly matches the reference. For the same mesh, we show in Fig.~\ref{signal_loss_reconstruction} the dMRI signal loss over the gradient descent iterations, which converges by iteration 250. The error in the reconstructed mesh converges around iteration 700 (Appendix Fig.~\ref{spatial_loss_reconstruction}) while the mesh volume converges around iteration 400 (Appendix Fig.~\ref{volume_change_over_iterations}). The latents in the SAE converge around iterations 400-600 (Appendix Fig.~\ref{Latent_space_over_iterations}).

\subsection{Ablation studies}

To systematically evaluate ReMiDi's ability to reconstruct meshes in various conditions, we performed several ablation studies on meshes with (1) bending and twisting, (2) beading and (3) fanning. In addition, we also ablated on (4) the number of diffusion directions and (5) diffusion times.

\subsubsection{Bending and Twisting}

\begin{figure}[ht]
\vskip 0.2in
\begin{center}
\centerline{\includegraphics[width=\columnwidth]{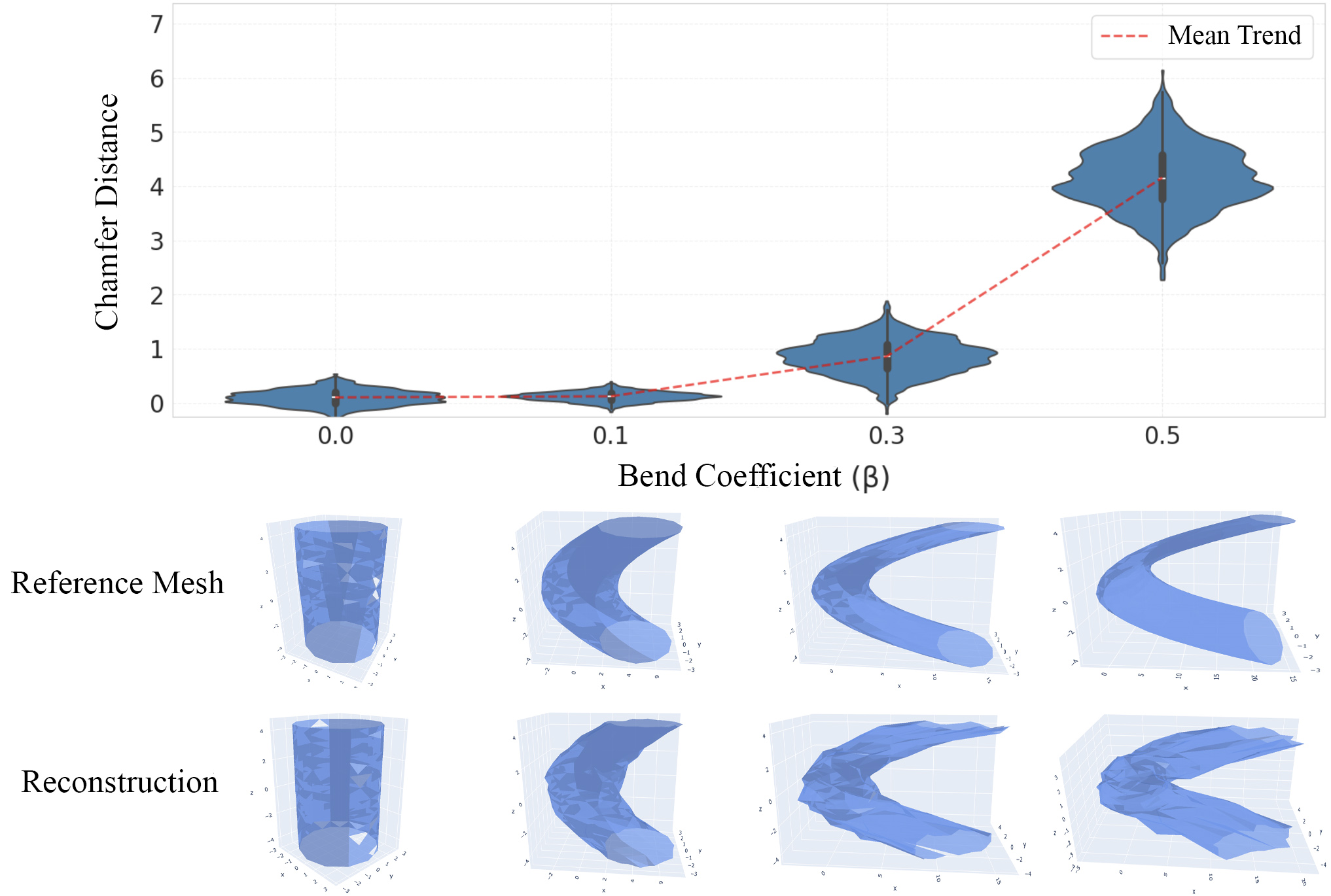}}
\caption{Reconstruction error as modified Chamfer distance between reference meshes (middle) and reconstructed meshes (bottom), for increasing bending in an axon.}
\label{Bending_reconstructions}
\end{center}
\vskip -0.2in
\end{figure}

For the bending and twisting dataset, we utilized a total of 35,136 meshes, divided into 26,136 training meshes, 4,500 validation meshes, and 4,500 testing meshes. The bending coefficient $(\beta)$ controls the deformation intensity, where points experience horizontal displacement proportional to their squared height $(h^2)$, with points at the base remaining fixed while those at the top undergo maximum displacement. The reconstructions were tested on twist values of 0.0 and 0.5, with bend values of 0.0, 0.1, 0.3, and 0.5, where $\beta=0.5$ produces 5 times greater deformation than $\beta=0.1$. Experiments were conducted over 750 iterations, using 3 sequences with different diffusion times and 30 directions. The findings from this investigation are depicted in Figure~\ref{Bending_reconstructions}. ReMiDi shows stable reconstructions as the bending coefficient increases, however there are some geometric irregularities obtained for higher bend coefficients.

\subsubsection{Beading}
\begin{figure}[ht]
\vskip 0.2in
\begin{center}
\centerline{\includegraphics[width=\columnwidth]{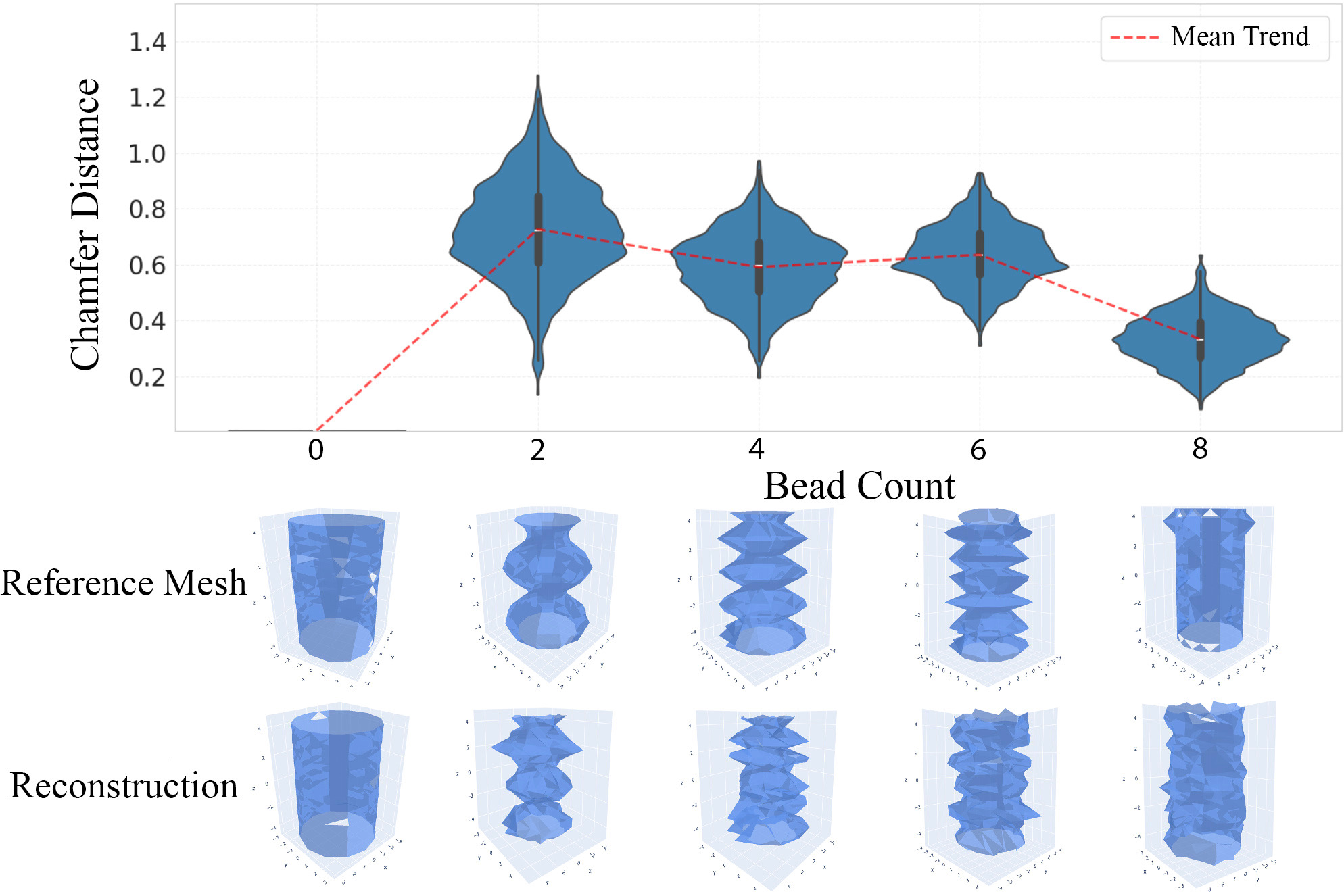}}
\caption{Reconstruction error as modified Chamfer distance between reference meshes (middle) and reconstructed meshes (bottom) for increasing number of beads along length of axon.}
\label{Beading_reconstructions}
\end{center}
\vskip -0.2in
\end{figure}

The beading dataset consisted of 33,696 meshes, divided into 24,696 training meshes and 4,500 meshes each for validation and testing. The training set exclusively featured axons with an odd number of beads, while the validation and testing sets comprised axons with an even number of beads. We tested on axons with 0, 2, 4, 6, and 8 beads, with bead sizes controlled using parameters of 0.3 and 0.4. Experiments were conducted over 750 iterations, utilizing 3 sequences and 30 directions. Figure ~\ref{Beading_reconstructions} shows that ReMiDi can reliably capture beading of axonal meshes. We note that for 8 beads or more, our 315-vertex mesh did not have enough resolution to properly capture the beads, hence the relatively flat surface. 

\subsubsection{Fanning}
\begin{figure}[ht]
\vskip 0.2in
\begin{center}
\centerline{\includegraphics[width=\columnwidth]{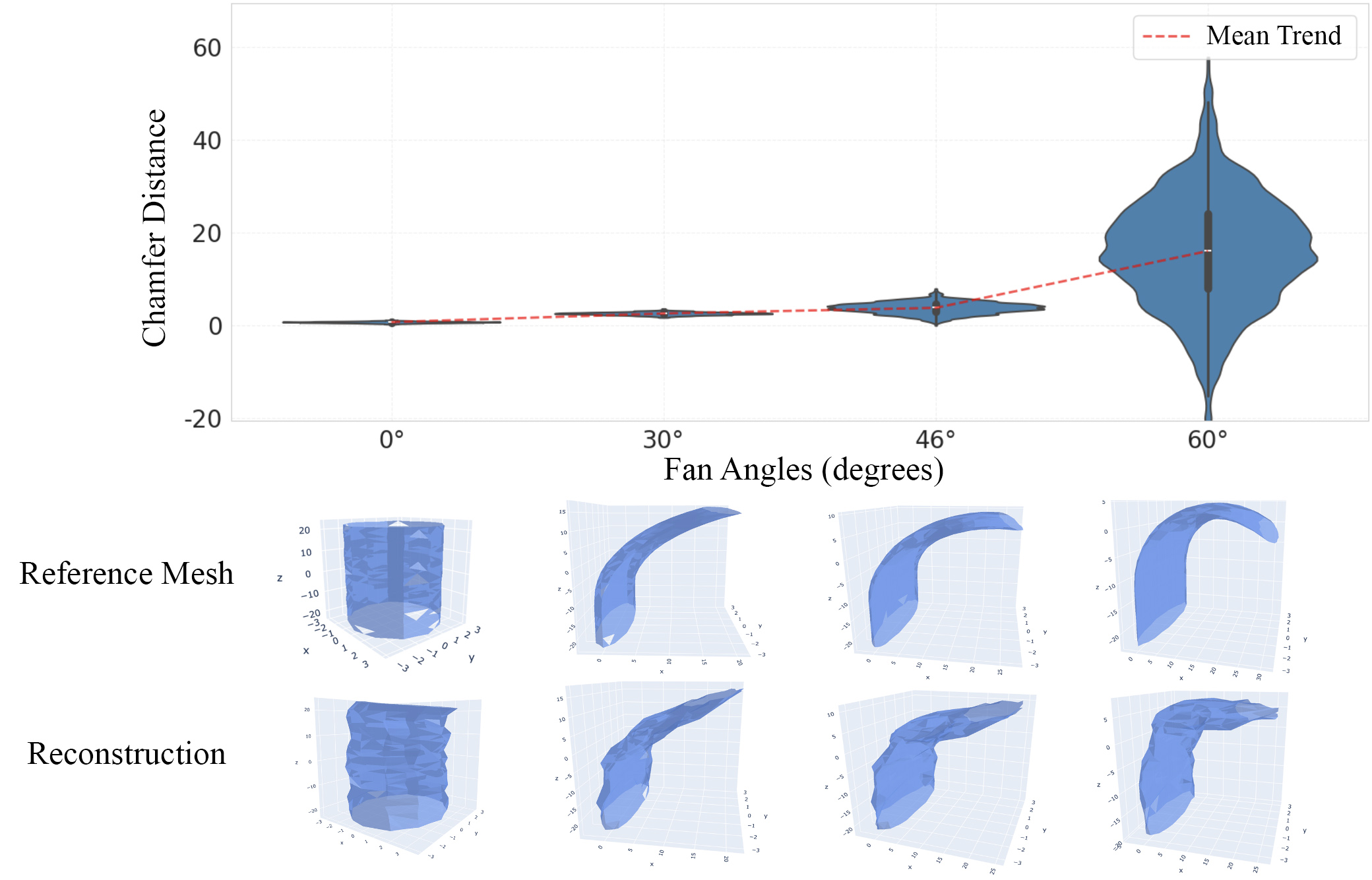}}
\caption{Modified Chamfer distance as error between reference meshes (middle) and reconstructed meshes (bottom) for increasing fan angles. While Chamfer distances are always non-negative, the visualization shows negative values at 60$^\circ$ due to applying KDE-based smoothing.}
\label{Fanning_reconstructions}
\end{center}
\vskip -0.2in
\end{figure}

For fanning deformations, we utilized a dataset of 35,136 meshes, divided into 26,136 training meshes and 4,500 meshes each for validation and testing. The dataset was carefully structured to ensure disjoint subsets: odd-numbered fanning angles were included in the training set, while even-numbered ones were allocated to the validation and testing sets. We test reconstructions on fanning angles of $0^\circ, 32^\circ, 46^\circ, \text{and } 60^\circ$. Experiments were conducted over 1000 iterations, using 3 sequences (different diffusion times) and 30 directions. The results of this experiment (Figure~\ref{Fanning_reconstructions}) show that ReMiDi can capture fanning to some extent, although the curvature of reconstructed meshes is not smooth and there are significant errors for 60$^\circ$ fanning angles.

\subsubsection{Diffusion directions and diffusion times}


\begin{figure}[ht]
\vskip 0.2in
\begin{center}
\centerline{\includegraphics[width=\columnwidth]{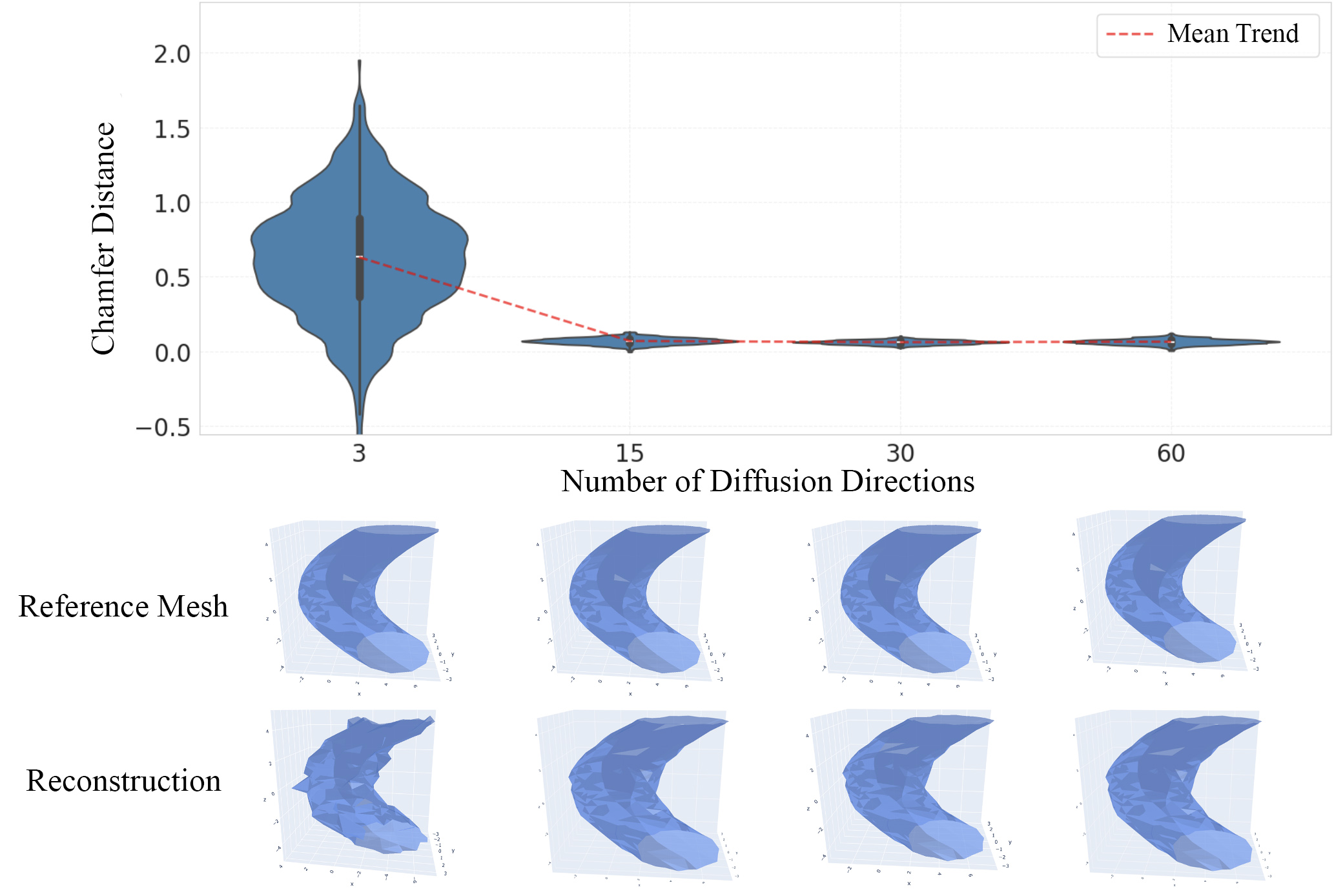}}
\caption{Modified Chamfer distance as error between reference meshes (middle) and reconstructed meshes (bottom) for increasing number of diffusion directions.}
\label{Directions_ablation}
\end{center}
\vskip -0.2in
\end{figure}


Diffusion directions specify the axes along which diffusion is measured within a voxel. To investigate the impact of the number of diffusion directions in the dMRI signal on reconstruction quality, we performed reconstructions of a bent axon using 3, 15, 30, and 60 directions, while using 3 sequences. Results in Fig.~\ref{Directions_ablation} show that ReMiDi can do some reconstructions even with as low as 3 diffusion directions, although having 15+ directions is ideal and it leads to smoother shapes. 

\begin{figure}[ht]
\vskip 0.2in
\begin{center}
\centerline{\includegraphics[width=\columnwidth]{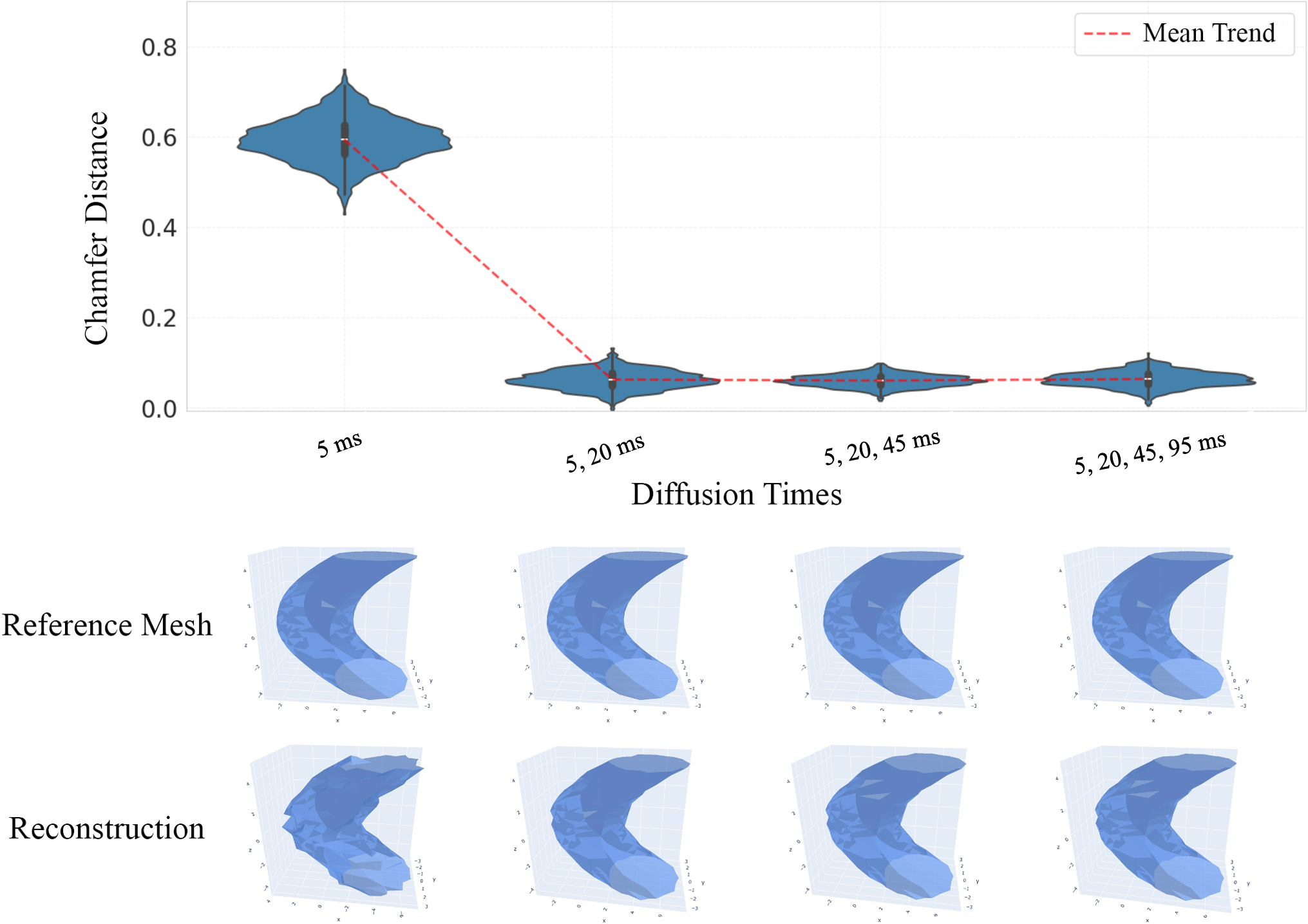}}
\caption{Modified Chamfer distance between reference meshes (middle) and reconstructed meshes (bottom) for increasing number of diffusion times in scan.}
\label{Diffusion_times_ablation}
\end{center}
\vskip -0.2in
\end{figure}


Multiple diffusion sequences with varying diffusion times are sensitive to varying amounts of displacements of water molecules -- long diffusion times capture slow motions while short diffusion times capture fast motions (see Fig.~\ref{SpinDoctor_vs_ReMiDi}). We conducted reconstructions of a bent axon over multiple diffusion sequences with varying combinations of diffusion times (5$ms$, 20$ms$, 45$ms$, 95$ms$) for each experiment, while using 30 diffusion directions. The results of this ablation (Fig.~\ref{Diffusion_times_ablation}) show significantly improved reconstruction when at least two sequences are used, with diffusion times of 5$ms$ and 20$ms$. 

\subsection{Comparison with a Neural Network}

To assess the reconstruction results of our work with a baseline, we compare the average reconstruction error between ReMiDi and a standard neural network (a multi-layered perceptron (MLP) \cite{rosenblattPerceptronProbabilisticModel1958} in this case) trained to reconstruct a plausible mesh for a given input signal. The MLP consists of 15 fully connected layers with 4096 hidden units in each layer. Dropout regularization is applied between layers to prevent overfitting, and the network is trained using mean squared error (MSE) loss. Table~\ref{tab:ReMiDi_vs_MLP} shows the error in reconstructed meshes between of ReMiDi against the MLP for reconstructing neuronal meshes from dMRI signals. We observe that ReMiDi achieves significantly lower spatial reconstruction error than the MLP in bending \& twisting and beading, which we attribute to ReMiDi physics-based inductive bias inherent in the simulation. However, in case of fanning, since the deformed mesh is much closer to the non-deformed mesh, the MLP is able to predict vertex positions with higher accuracy. Conversely, for reconstructions involving significant changes in vertex positions, ReMiDi outperforms the MLP.

\begin{table}[t]
    \centering
    \caption{Comparison between reconstructed meshes by ReMiDi and an MLP. Numbers represent the modified Chamfer distance.}
    \vskip 0.15in
    \label{tab:ReMiDi_vs_MLP}
    \begin{tabular}{@{}lcc@{}}  
    \toprule[1.5pt]  
    \textbf{Deformation} & \textbf{ReMiDi} & \textbf{MLP} \\
    \midrule
    Bending \& Twisting & \textcolor{green!70!black}{1.709 $\pm$ 0.261} & \textcolor{red!70!black}{43.73 $\pm$ 14.99} \\
    Beading & \textcolor{green!70!black}{0.568 $\pm$ 0.086} & \textcolor{red!70!black}{45.01 $\pm$ 5.13} \\
    Fanning & \textcolor{red!70!black}{7.433 $\pm$ 4.844} & \textcolor{green!70!black}{0.743 $\pm$ 0.24} \\
    \bottomrule[1.5pt]  
    \end{tabular}
    \vskip -0.1in
\end{table}


\section{Conclusion}

We introduce ReMiDi, a tool designed to reconstruct neuronal microstructure from dMRI signals. Our framework combines a SAE with a differentiable dMRI simulator, allowing iterative updates and reconstruction of the mesh. By leveraging the SAE and differentiable simulator, ReMiDi minimizes the difference between simulated and reference signals, resulting in plausible voxel microstructures. ReMiDi ensures that the generated microstructure mesh matches the reference signal. The accuracy of the shape in producing the correct signal can be verified through back-substitution, confirming the consistency of the reconstruction process. Currently, ReMiDi operates with synthetic data and simplified voxel structures, demonstrating a proof of concept for reconstructing neuronal features.

However, ReMiDi still faces several challenges. The current simulator has difficulty handling complex voxel geometries due to significant GPU memory requirements: the largest meshes we could fit on an NVIDIA A10 GPU contained around 1,000 vertices. Future improvements will focus on optimizing parallelization and GPU concurrency. Additionally, moving from synthetic data to real-world MRI scans presents challenges due to lower data quality,  high complexity of true microstructure (a single MRI voxel contains thousands of neurons) and ill-posedness. Additionally, our pipeline is currently only capable to optimize single-compartment meshes, although we next plan to extend it to multi-compartment. Addressing these limitations would push in-vivo brain microstructure mapping to the next level. 


\section*{Impact Statement}
Our work achieves a next-generation dMRI reconstruction, opening the ability to reconstruct brain microstructure with arbitrary meshes. This will enable in-vivo mesoscopic mapping of the human brain. We foresee significant applications towards improved biomarkers for many neurodegenerative diseases, including Alzheimer's disease, Multiple Sclerosis, Parkinson's, Traumatic Brain Injury, as well as mapping brain tumors. In addition, this work could bring diffusion tractography to the next level and enable significantly better connectomes of the human brain, with improves rates of false positives and false negatives of white-matter tracts. Diffusion MRI connectomes can potentially be used as starting points for whole brain simulations, likely through spiking neural network \cite{TAVANAEI201947} (SNN)-based simulations.

\nocite{langley00}

\bibliography{icml_final_paper}

\begin{thebibliography}{35}
\providecommand{\natexlab}[1]{#1}
\providecommand{\url}[1]{\texttt{#1}}
\expandafter\ifx\csname urlstyle\endcsname\relax
  \providecommand{\doi}[1]{doi: #1}\else
  \providecommand{\doi}{doi: \begingroup \urlstyle{rm}\Url}\fi

\bibitem[Barrow et~al.(1977)Barrow, Tenenbaum, Bolles, and Wolf]{Barrow1977}
Barrow, H.~G., Tenenbaum, J.~M., Bolles, R.~C., and Wolf, H.~C.
\newblock Parametric correspondence and chamfer matching: Two new techniques for image matching.
\newblock In \emph{Proceedings of the 5th International Joint Conference on Artificial Intelligence (IJCAI)}, pp.\  659--663, Cambridge, MA, 1977.
\newblock URL \url{https://www.ijcai.org/Proceedings/77-2/Papers/024.pdf}.

\bibitem[Basser et~al.(1994)Basser, Mattiello, and LeBihan]{Basser1994}
Basser, P.~J., Mattiello, J., and LeBihan, D.
\newblock Mr diffusion tensor spectroscopy and imaging.
\newblock \emph{Biophysical Journal}, 66\penalty0 (1):\penalty0 259--267, 1994.
\newblock \doi{10.1016/S0006-3495(94)80775-1}.
\newblock URL \url{https://doi.org/10.1016/S0006-3495(94)80775-1}.

\bibitem[Benitez et~al.(2014)Benitez, Fieremans, Jensen, Falangola, Tabesh, Ferris, and Helpern]{BENITEZ201464}
Benitez, A., Fieremans, E., Jensen, J.~H., Falangola, M.~F., Tabesh, A., Ferris, S.~H., and Helpern, J.~A.
\newblock White matter tract integrity metrics reflect the vulnerability of late-myelinating tracts in alzheimer's disease.
\newblock \emph{NeuroImage: Clinical}, 4:\penalty0 64--71, 2014.
\newblock ISSN 2213-1582.
\newblock \doi{https://doi.org/10.1016/j.nicl.2013.11.001}.
\newblock URL \url{https://www.sciencedirect.com/science/article/pii/S2213158213001496}.

\bibitem[Bihan et~al.(1986)Bihan, Breton, Lallemand, Grenier, Cabanis, and Laval-Jeantet]{LeBihan1986}
Bihan, D.~L., Breton, E., Lallemand, D., Grenier, P., Cabanis, E., and Laval-Jeantet, M.
\newblock Mr imaging of intravoxel incoherent motions: Application to diffusion and perfusion in neurologic disorders.
\newblock \emph{Radiology}, 161\penalty0 (2):\penalty0 401--407, 1986.
\newblock \doi{10.1148/radiology.161.2.3763909}.
\newblock URL \url{https://pubs.rsna.org/doi/abs/10.1148/radiology.161.2.3763909}.

\bibitem[Bihan et~al.(2001)Bihan, Mangin, Poupon, Clark, Pappata, Molko, and Chabriat]{LeBihan2001}
Bihan, D.~L., Mangin, J.-F., Poupon, C., Clark, C.~A., Pappata, S., Molko, N., and Chabriat, H.
\newblock Diffusion tensor imaging: concepts and applications.
\newblock \emph{Journal of Magnetic Resonance Imaging}, 13\penalty0 (4):\penalty0 534--546, April 2001.
\newblock \doi{10.1002/jmri.1076}.

\bibitem[Bodini et~al.(2009)Bodini, Khaleeli, Cercignani, Miller, Thompson, and Ciccarelli]{bodini2009exploring}
Bodini, B., Khaleeli, Z., Cercignani, M., Miller, D.~H., Thompson, A.~J., and Ciccarelli, O.
\newblock Exploring the relationship between white matter and gray matter damage in early primary progressive multiple sclerosis: an in vivo study with tbss and vbm.
\newblock \emph{Human Brain Mapping}, 30\penalty0 (9):\penalty0 2852--2861, 2009.
\newblock \doi{10.1002/hbm.20713}.
\newblock URL \url{https://pubmed.ncbi.nlm.nih.gov/19172648/}.

\bibitem[Boyd(2001)]{boyd01}
Boyd, J.~P.
\newblock \emph{{Chebyshev} and {Fourier} Spectral Methods}.
\newblock Dover Books on Mathematics. Dover Publications, Mineola, NY, second edition, 2001.
\newblock ISBN 0486411834 9780486411835.

\bibitem[Budde \& Frank(2010)Budde and Frank]{Budde2010}
Budde, M.~D. and Frank, J.~A.
\newblock Neurite beading is sufficient to decrease the apparent diffusion coefficient after ischemic stroke.
\newblock \emph{Proceedings of the National Academy of Sciences}, 107\penalty0 (32):\penalty0 14472--14477, 2010.
\newblock \doi{10.1073/pnas.1004841107}.
\newblock URL \url{https://www.pnas.org/content/107/32/14472}.

\bibitem[Callaghan(1997)]{CALLAGHAN199774}
Callaghan, P.~T.
\newblock A simple matrix formalism for spin echo analysis of restricted diffusion under generalized gradient waveforms.
\newblock \emph{Journal of Magnetic Resonance}, 129\penalty0 (1):\penalty0 74--84, 1997.
\newblock ISSN 1090-7807.
\newblock \doi{https://doi.org/10.1006/jmre.1997.1233}.
\newblock URL \url{https://www.sciencedirect.com/science/article/pii/S1090780797912337}.

\bibitem[Callaghan et~al.(2020)Callaghan, Alexander, Palombo, and Zhang]{CALLAGHAN2020117107}
Callaghan, R., Alexander, D.~C., Palombo, M., and Zhang, H.
\newblock Config: Contextual fibre growth to generate realistic axonal packing for diffusion mri simulation.
\newblock \emph{NeuroImage}, 220:\penalty0 117107, 2020.
\newblock ISSN 1053-8119.
\newblock \doi{https://doi.org/10.1016/j.neuroimage.2020.117107}.
\newblock URL \url{https://www.sciencedirect.com/science/article/pii/S1053811920305930}.

\bibitem[Cook et~al.(2006)Cook, Bai, Nedjati-Gilani, Seunarine, Hall, Parker, and Alexander]{Cook2006Camino}
Cook, P.~A., Bai, Y., Nedjati-Gilani, S., Seunarine, K.~K., Hall, M.~G., Parker, G.~J., and Alexander, D.~C.
\newblock Camino: Open-source diffusion-mri reconstruction and processing.
\newblock In \emph{Proceedings of the 14th Annual Meeting of ISMRM}, pp.\  2759, Seattle, WA, USA, 2006.
\newblock URL \url{https://cds.ismrm.org/protected/06MProceedings/PDFfiles/02759.pdf}.

\bibitem[Ho-Le(1988)]{HOLE198827}
Ho-Le, K.
\newblock Finite element mesh generation methods: a review and classification.
\newblock \emph{Computer-Aided Design}, 20\penalty0 (1):\penalty0 27--38, 1988.
\newblock ISSN 0010-4485.
\newblock \doi{https://doi.org/10.1016/0010-4485(88)90138-8}.
\newblock URL \url{https://www.sciencedirect.com/science/article/pii/0010448588901388}.

\bibitem[Hoang \& Schwab(2005)Hoang and Schwab]{Hoang2005}
Hoang, V.~H. and Schwab, C.
\newblock High-dimensional finite elements for elliptic problems with multiple scales.
\newblock \emph{Multiscale Modeling \& Simulation}, 3\penalty0 (1):\penalty0 168--194, 2005.
\newblock \doi{10.1137/030601077}.
\newblock URL \url{https://doi.org/10.1137/030601077}.

\bibitem[Ianus et~al.(2021)Ianus, Alexander, Zhang, and Palombo]{Ianus2021}
Ianus, A., Alexander, D.~C., Zhang, H., and Palombo, M.
\newblock Mapping complex cell morphology in the grey matter with double diffusion encoding mr: A simulation study.
\newblock \emph{NeuroImage}, 241:\penalty0 118424, 2021.
\newblock \doi{10.1016/j.neuroimage.2021.118424}.
\newblock URL \url{https://doi.org/10.1016/j.neuroimage.2021.118424}.

\bibitem[Jeurissen et~al.(2019)Jeurissen, Descoteaux, Mori, and Leemans]{Jeurissen2019Diffusion}
Jeurissen, B., Descoteaux, M., Mori, S., and Leemans, A.
\newblock Diffusion mri fiber tractography of the brain.
\newblock \emph{NMR in Biomedicine}, 32\penalty0 (4):\penalty0 e3785, 2019.
\newblock \doi{10.1002/nbm.3785}.
\newblock e3785 NBM-17-0045.R2.

\bibitem[Johnson(2009)]{Johnson2009}
Johnson, C.
\newblock \emph{Numerical Solution of Partial Differential Equations by the Finite Element Method}.
\newblock Dover Publications, New York, 1st edition, 2009.
\newblock ISBN 9780486469003.

\bibitem[Kasim \& Vinko(2020)Kasim and Vinko]{Kasim2020xiTorch}
Kasim, M.~F. and Vinko, S.~M.
\newblock $\xi$-torch: differentiable scientific computing library.
\newblock \emph{arXiv preprint arXiv:2010.01921}, 2020.
\newblock URL \url{https://arxiv.org/abs/2010.01921}.

\bibitem[Kerkelä et~al.(2020)Kerkelä, Nery, Hall, and Clark]{Kerkelä2020}
Kerkelä, L., Nery, F., Hall, M.~G., and Clark, C.~A.
\newblock Disimpy: A massively parallel monte carlo simulator for generating diffusion-weighted mri data in python.
\newblock \emph{Journal of Open Source Software}, 5\penalty0 (52):\penalty0 2527, 2020.
\newblock \doi{10.21105/joss.02527}.
\newblock URL \url{https://doi.org/10.21105/joss.02527}.

\bibitem[Langley(2000)]{langley00}
Langley, P.
\newblock Crafting papers on machine learning.
\newblock In Langley, P. (ed.), \emph{Proceedings of the 17th International Conference on Machine Learning (ICML 2000)}, pp.\  1207--1216, Stanford, CA, 2000. Morgan Kaufmann.

\bibitem[Lemeunier et~al.(2022)Lemeunier, Denis, Lavoué, and Dupont]{LEMEUNIER2022131}
Lemeunier, C., Denis, F., Lavoué, G., and Dupont, F.
\newblock Representation learning of 3d meshes using an autoencoder in the spectral domain.
\newblock \emph{Computers \& Graphics}, 107:\penalty0 131--143, 2022.
\newblock ISSN 0097-8493.
\newblock \doi{https://doi.org/10.1016/j.cag.2022.07.011}.
\newblock URL \url{https://www.sciencedirect.com/science/article/pii/S0097849322001285}.

\bibitem[Li et~al.(2019)Li, Nguyen, Tran, Valdman, Trang, Nguyen, Vu, Tran, Tran, and Nguyen]{Li2019SpinDoctor}
Li, J.-R., Nguyen, V.-D., Tran, T.~N., Valdman, J., Trang, C.-B., Nguyen, K.~V., Vu, D. T.~S., Tran, H.~A., Tran, H. T.~A., and Nguyen, T. M.~P.
\newblock Spindoctor: a matlab toolbox for diffusion mri simulation.
\newblock \emph{arXiv preprint arXiv:1902.01025}, 2019.
\newblock URL \url{https://arxiv.org/abs/1902.01025}.

\bibitem[Logan(2017)]{Logan2017}
Logan, D.~L.
\newblock \emph{A First Course in the Finite Element Method}.
\newblock Cengage Learning, 6th edition, 2017.
\newblock ISBN 9780357704745.

\bibitem[Palombo et~al.(2020)Palombo, Ianus, Guerreri, Nunes, Alexander, Shemesh, and Zhang]{PALOMBO2020116835}
Palombo, M., Ianus, A., Guerreri, M., Nunes, D., Alexander, D.~C., Shemesh, N., and Zhang, H.
\newblock Sandi: A compartment-based model for non-invasive apparent soma and neurite imaging by diffusion mri.
\newblock \emph{NeuroImage}, 215:\penalty0 116835, 2020.
\newblock ISSN 1053-8119.
\newblock \doi{https://doi.org/10.1016/j.neuroimage.2020.116835}.
\newblock URL \url{https://www.sciencedirect.com/science/article/pii/S1053811920303220}.

\bibitem[Palombo et~al.(2022)Palombo, Zhang, and Alexander]{Palombo2022}
Palombo, M., Zhang, H., and Alexander, D.~C.
\newblock A generative model of realistic brain cells with application to numerical simulation of the diffusion-weighted mr signal.
\newblock \emph{NeuroImage}, 245:\penalty0 118424, 2022.
\newblock \doi{10.1016/j.neuroimage.2021.118424}.
\newblock URL \url{https://doi.org/10.1016/j.neuroimage.2021.118424}.

\bibitem[Paszke et~al.(2019)Paszke, Gross, Massa, Lerer, Bradbury, Chanan, Killeen, Lin, Gimelshein, Antiga, Desmaison, Köpf, Yang, DeVito, Raison, Tejani, Chilamkurthy, Steiner, Fang, Bai, and Chintala]{Paszke2019}
Paszke, A., Gross, S., Massa, F., Lerer, A., Bradbury, J., Chanan, G., Killeen, T., Lin, Z., Gimelshein, N., Antiga, L., Desmaison, A., Köpf, A., Yang, E., DeVito, Z., Raison, M., Tejani, A., Chilamkurthy, S., Steiner, B., Fang, L., Bai, J., and Chintala, S.
\newblock Pytorch: an imperative style, high-performance deep learning library.
\newblock \emph{Advances in Neural Information Processing Systems}, 32:\penalty0 8024--8035, 2019.

\bibitem[Reddy(2006)]{Reddy2006}
Reddy, J.~N.
\newblock \emph{Introduction to the Finite Element Method}.
\newblock McGraw-Hill Education, New York, 3rd edition, 2006.
\newblock ISBN 9780072466850.
\newblock URL \url{https://www.accessengineeringlibrary.com/content/book/9780072466850}.

\bibitem[Rosenblatt(1958)]{rosenblattPerceptronProbabilisticModel1958}
Rosenblatt, F.
\newblock The perceptron: {{A}} probabilistic model for information storage and organization in the brain.
\newblock \emph{Psychological Review}, 65\penalty0 (6):\penalty0 386--408, 1958.
\newblock ISSN 1939-1471, 0033-295X.
\newblock \doi{10.1037/h0042519}.

\bibitem[Sotiropoulos et~al.(2012)Sotiropoulos, Behrens, and Jbabdi]{Sotiropoulos2012}
Sotiropoulos, S.~N., Behrens, T.~E., and Jbabdi, S.
\newblock Ball and rackets: Inferring fibre fanning from diffusion-weighted mri.
\newblock \emph{NeuroImage}, 60\penalty0 (2):\penalty0 1412--1425, 2012.
\newblock \doi{10.1016/j.neuroimage.2012.01.056}.
\newblock URL \url{https://www.ncbi.nlm.nih.gov/pmc/articles/PMC3304013/}.

\bibitem[Stejskal \& Tanner(1965)Stejskal and Tanner]{Stejskal1965}
Stejskal, E.~O. and Tanner, J.~E.
\newblock Spin diffusion measurements: Spin echoes in the presence of a time‐dependent field gradient.
\newblock \emph{The Journal of Chemical Physics}, 42\penalty0 (1):\penalty0 288--292, 1965.
\newblock \doi{10.1063/1.1695690}.
\newblock URL \url{https://doi.org/10.1063/1.1695690}.

\bibitem[Tavanaei et~al.(2019)Tavanaei, Ghodrati, Kheradpisheh, Masquelier, and Maida]{TAVANAEI201947}
Tavanaei, A., Ghodrati, M., Kheradpisheh, S.~R., Masquelier, T., and Maida, A.
\newblock Deep learning in spiking neural networks.
\newblock \emph{Neural Networks}, 111:\penalty0 47--63, 2019.
\newblock ISSN 0893-6080.
\newblock \doi{https://doi.org/10.1016/j.neunet.2018.12.002}.
\newblock URL \url{https://www.sciencedirect.com/science/article/pii/S0893608018303332}.

\bibitem[Torrey(1956)]{PhysRev.104.563}
Torrey, H.~C.
\newblock Bloch equations with diffusion terms.
\newblock \emph{Phys. Rev.}, 104:\penalty0 563--565, Nov 1956.
\newblock \doi{10.1103/PhysRev.104.563}.
\newblock URL \url{https://link.aps.org/doi/10.1103/PhysRev.104.563}.

\bibitem[Turner et~al.(1956)Turner, Clough, Martin, and Topp]{Turner1956}
Turner, M.~J., Clough, R.~W., Martin, H.~C., and Topp, L.~J.
\newblock Stiffness and deflection analysis of complex structures.
\newblock \emph{Journal of the Aeronautical Sciences}, 23\penalty0 (9):\penalty0 805--823, 1956.
\newblock \doi{10.2514/8.3664}.
\newblock URL \url{https://arc.aiaa.org/doi/abs/10.2514/8.3664}.

\bibitem[{University of California, San Diego}(2021)]{UCSD2021}
{University of California, San Diego}.
\newblock Neuronal axons optimally balance speed and processing time.
\newblock \emph{UC San Diego Jacobs School of Engineering News}, February 2021.
\newblock URL \url{https://jacobsschool.ucsd.edu/news/release/2580?id=2580}.
\newblock Accessed: January 28, 2025.

\bibitem[Zhang et~al.(2012)Zhang, Schneider, Wheeler-Kingshott, and Alexander]{zhang2012noddi}
Zhang, H., Schneider, T., Wheeler-Kingshott, C.~A., and Alexander, D.~C.
\newblock Noddi: practical in vivo neurite orientation dispersion and density imaging of the human brain.
\newblock \emph{NeuroImage}, 61\penalty0 (4):\penalty0 1000--1016, 2012.
\newblock \doi{10.1016/j.neuroimage.2012.03.072}.
\newblock URL \url{https://pubmed.ncbi.nlm.nih.gov/22484410/}.

\bibitem[Zhao et~al.(2022)Zhao, Lindell, and Wetzstein]{Zhao2022}
Zhao, Q., Lindell, D.~B., and Wetzstein, G.
\newblock Learning to solve pde-constrained inverse problems with graph networks.
\newblock \emph{arXiv preprint arXiv:2206.00711}, 2022.
\newblock URL \url{https://arxiv.org/abs/2206.00711}.

\end{thebibliography}
\bibliographystyle{icml2025}

\newpage
\appendix
\onecolumn



\section{Deformations of Axons}


\par{A single voxel in an MRI scan (e.g. 1x1x1 mm) can contain thousands of neuronal fibers, which all contribute to the measured diffusion MRI signal. Recovering the full distribution of fiber shapes and orientations poses a highly ill-posed problem \cite{Jeurissen2019Diffusion}. }


\begin{figure}[ht]
\vskip 0.2in
\begin{center}
\centerline{\includegraphics[width=0.6\columnwidth]{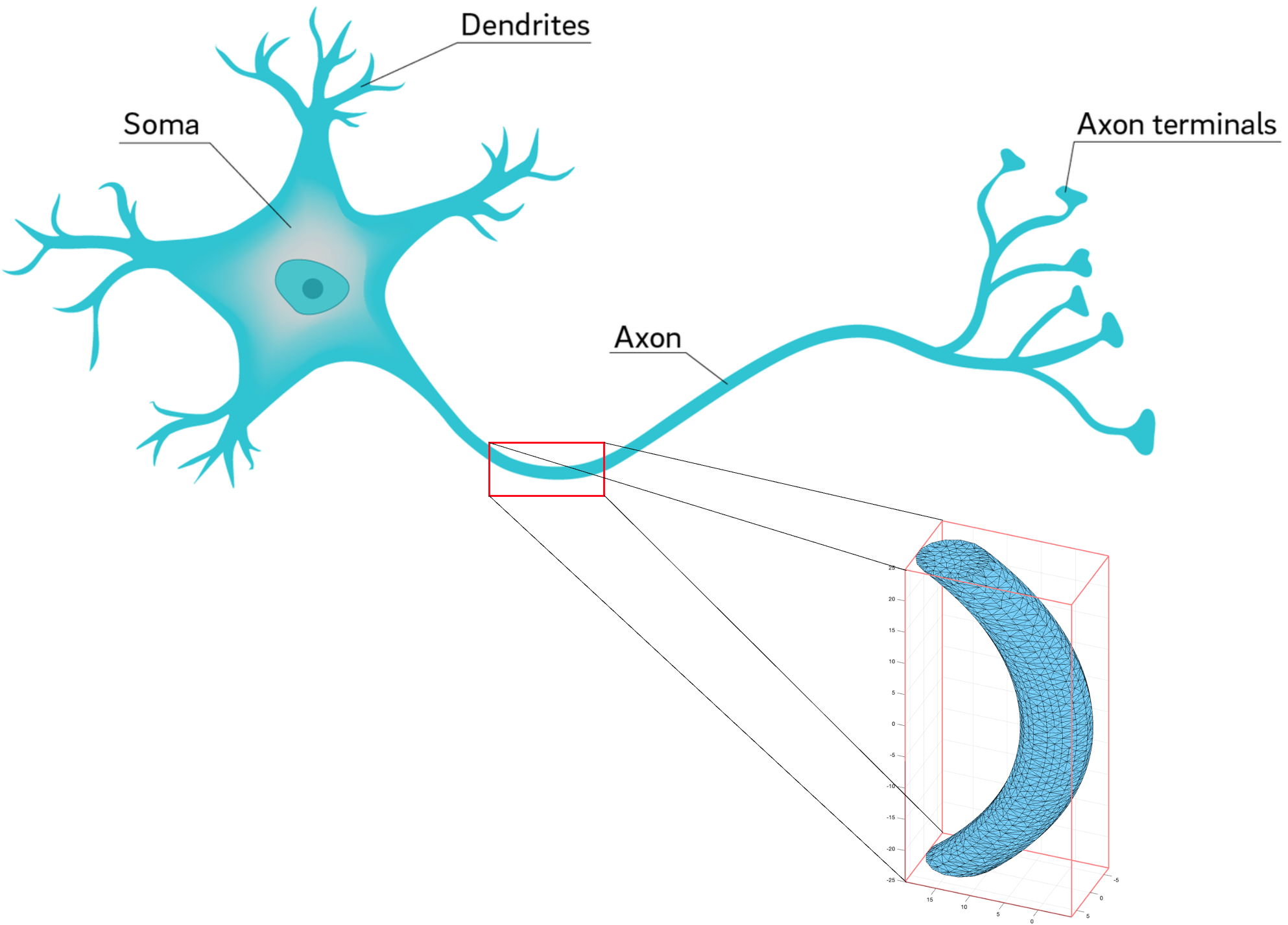}}
\caption{Cylindrical section of a bent region of neuron \cite{UCSD2021}}
\label{Cylinder_section_neuron}
\end{center}
\vskip -0.2in
\end{figure}

\par{In the brain, the neurons are connected to each other by a thin elongated process called the axon. Axons transmit electrical signals from the neuron to the dendrites of other neurons, which in turn relay the signals back to their respective neuron bodies.}

\par{As axons grow, they develop an elongated and curved shape that facilitates the efficient transmission of electrical signals between neurons \cite{UCSD2021}. However, they do not always resemble thin, elongated cylinders. Instead, they may exhibit various well-known structural deformations, including but not limited to bending and twisting, beading, and fanning.}

\par{The bends and twists observed in axons as they elongate and grow are collectively referred to as bending in our work. This phenomenon typically arises from various mechanical forces acting on the external surface of axons. These forces include the flow of fluids, such as blood and cerebrospinal fluid, as well as mechanical compression resulting from interactions with nearby neurons.}

\par{Fanning is a type of deformation similar to bending; however, in the case of fanning, axons exhibit only slight bending without any tortuosity around their vertical axis. This deformation can be described as the observed convergence of axons as they approach the thin outer membrane of other neurons, known as the cortex, or their divergence as they move away from it. The differences in fanning behavior are often attributed to the polarity of axons. However, in our work, we do not account for polarity and instead represent fanning simply as a divergence from the vertical axis of the axons \cite{Sotiropoulos2012}}.

\par{One of the most intriguing deformations we address is beading. Beading refers to the local enlargements and constrictions observed along axons, often resulting from acute brain injuries. These deformations are readily detectable through dMRI, as they cause a distinct change in the Apparent Diffusion Coefficient (ADC). Beading can be modeled as a series of consecutive beads along the axon's length. In our work, we assume these beads are of uniform size and represent them by applying sine or cosine transformations along the surface of the axon \cite{Budde2010}}.

\par{In our work, we examine these three types of deformations in a single cylindrical axon. However, our primary focus is on a small cylindrical section of the axon as shown in figure ~\ref{Cylinder_section_neuron}, where such deformations are more pronounced and easier to model using tetrahedral meshes.}

\section{Finite Element Method and Finite Element Mesh}

\par{Finite Element Method is a set of techniques used for numerically solving PDEs involving flow of fluids, transfer of heat, and, in our case, to model the evolution of magnetic flux throughout a voxel given a set of well defined boundary and interface conditions. 

To solve a problem, the FEM breaks it down into a set of simpler, smaller components known as finite elements (FEs). This process involves spatial discretization, accomplished by constructing a mesh that represents the object, commonly referred to as the FEM of the object. This approach transforms the problem into a finite numerical domain, where the solution has a finite number of points. Then, at each discrete point, the unknown quantities are approximated using basis functions within each FE. These local approximations are systematically assembled into a global system of algebraic equations that represent the entire problem domain. By applying numerical solvers to this system, it yields an approximation of the solution within the defined boundary conditions.}

\par{We make use of similar finite element meshes to represent the singular axons, and their corresponding deformations such as bending, fanning and beading within a voxel. These meshes are 3D volumetric tetrahedral meshes each having 315 vertices.}

\par{These meshes are generated by Spindoctor Toolbox which internally calls Tetgen to generate these tetrahedral meshes.} Figure~\ref{Cylinder_section_neuron} shows how the meshes are used to represent a section of interest in the axonal fibres. The mesh is made up of smaller tetrahedra, along with helping model it numerically are also helpful in representing complex shapes \cite{Hoang2005,Reddy2006,Logan2017}.

\section{Diffusion MRI}
Diffusion MRI is a non-invasive imaging technique that evaluates the microstructural properties of tissues by analyzing the motion of water molecules. Using diffusion-weighted imaging (DWI), it applies magnetic field gradients to measure water displacement. Signal attenuation in DWI is modeled by the Stejskal–Tanner equation\cite{Stejskal1965}: 
$$S(b) = S_0 \cdot e^{-b \cdot D}$$
where \( b \) is the diffusion weighting factor determined by the strength and timing of the gradient, \( S(b) \) is the intensity of the signal with diffusion weighting $b$, \( S_0 \) is the signal without weighting, and \( D \) is the apparent diffusion coefficient (ADC). ADC quantifies water diffusion within tissues, with lower values indicating restricted diffusion and higher values free diffusion \cite{Stejskal1965}. In diffusion tensor imaging (DTI), a \( 3 \times 3 \) diffusion tensor describes directional water diffusivity, and eigenvalue decomposition of the tensor yields metrics like fractional anisotropy and mean diffusivity, which provide insights into tissue structure \cite{LeBihan2001}.

\section{Spectral Auto-Encoder}

The spectral domain offers significant advantages for processing meshes with consistent connectivity, as the Laplacian operator provides a shared spectral basis for all meshes in a dataset. This shared basis eliminates the need to recompute spatial relationships for each mesh, facilitating efficient handling of large datasets.

Spectral Auto-Encoders (SAEs) \cite{LEMEUNIER2022131} operate exclusively in the spectral domain, in contrast to Chebyshev methods \cite{boyd01}, which switch between the spectral and spatial domains, resulting in higher computational overhead. SAEs focus on a truncated set of spectral coefficients, primarily the low-frequency components, which reduces the input size and accelerates training and inference. This emphasis on low-frequency components ensures smoother and more realistic reconstructions by capturing the dominant features of the geometry.

The graph Laplacian $L$ of a 3D mesh, defined over a set of points $P$, is computed as:
\[L = D - A,\]
where \( A \) is the adjacency matrix representing vertex connectivity, and \( D \) is the diagonal degree matrix containing the degree of each vertex. The eigenvectors of 
$L$ are computed by solving the eigenvalue problem:
\[L\Phi = \Lambda\Phi,\]
where \( \Lambda \) is a diagonal matrix of eigenvalues, and \( \Phi \) is the matrix of eigenvectors. These eigenvectors allow us to transform the vertex coordinates $P$ into the spectral coefficients $C$ using:
\[C = \Phi^T P,\]
and the inverse transformation from the spectral coefficients to the vertex coordinates is given by: 
$$P = \Phi C.$$
The graph Laplacian is analogous to the Fourier basis in the spectral domain, with its eigenvectors representing \lq frequencies\rq. When mesh geometry is projected onto these eigenvectors to compute spectral coefficients, the magnitude of the coefficients indicates the amount of the geometry's energy at a given frequency. Low frequencies generally capture the coarse shape (where most of the energy is concentrated), while high frequencies represent finer geometric details. For datasets with consistent connectivity, the eigenvectors are computed once from a representative mesh and then reused as a common spectral basis for all meshes. This consistency ensures reliable and efficient spectral transformations across the dataset.

\section{More Details of Reconstructions by ReMiDi}

\subsection{Spatial loss during reconstruction}

\begin{figure}[ht]
\vskip 0.2in
\begin{center}
\centerline{\includegraphics[width=0.5\columnwidth]{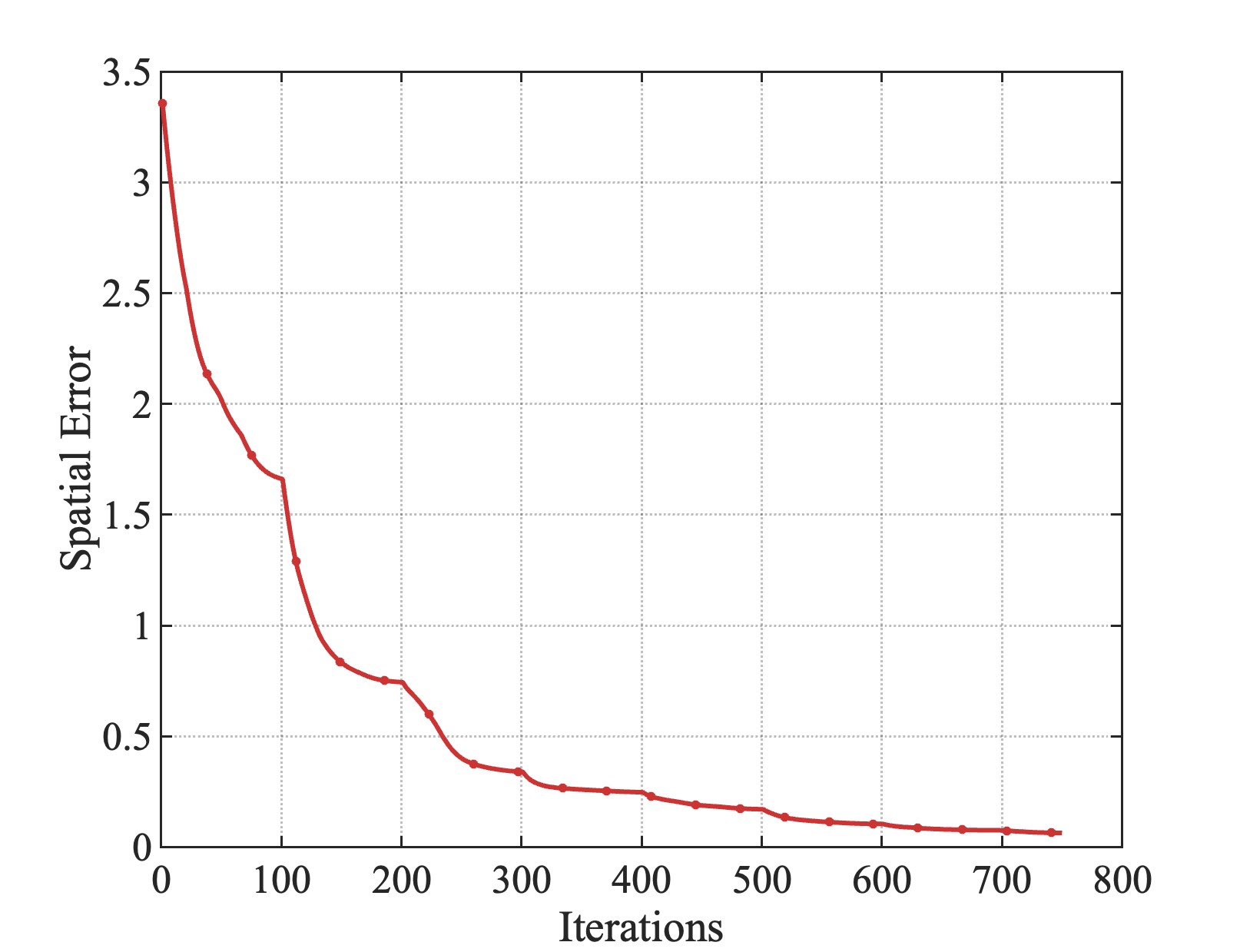}}
\caption{Chamfer distance (measuring spatial error) with respect to the reference mesh over gradient descent iterations, while reconstructing a bent axon using ReMiDi}
\label{spatial_loss_reconstruction}
\end{center}
\vskip -0.2in
\end{figure}

The above figure~\ref{spatial_loss_reconstruction} shows the change in spatial error over the iterative reconstruction process by ReMiDi, the spatial error here is computed using Chamfer Distance between the mesh at each iteration and the reference mesh.

Figures~\ref{Iterative_reconstruction_steps},~\ref{signal_loss_reconstruction}, and~\ref{spatial_loss_reconstruction} provide a visual and quantitative overview of an iterative reconstruction by ReMiDi. Figures~\ref{signal_loss_reconstruction}, and~\ref{spatial_loss_reconstruction} highlight the ill-posed nature of the inverse reconstruction problem. Specifically, we observe a steep drop in signal loss between iteration 1 and 100, but the same is not true for spatial error. Additionally, while the signal loss converges around iteration 250, the spatial loss takes significantly longer to converge. This highlights the inherent ill-posedness of the reconstruction problem.

\subsection{Volume Change during Iterative Reconstruction}

\begin{figure}[ht]
\vskip 0.2in
\begin{center}
\centerline{\includegraphics[width=0.5\columnwidth]{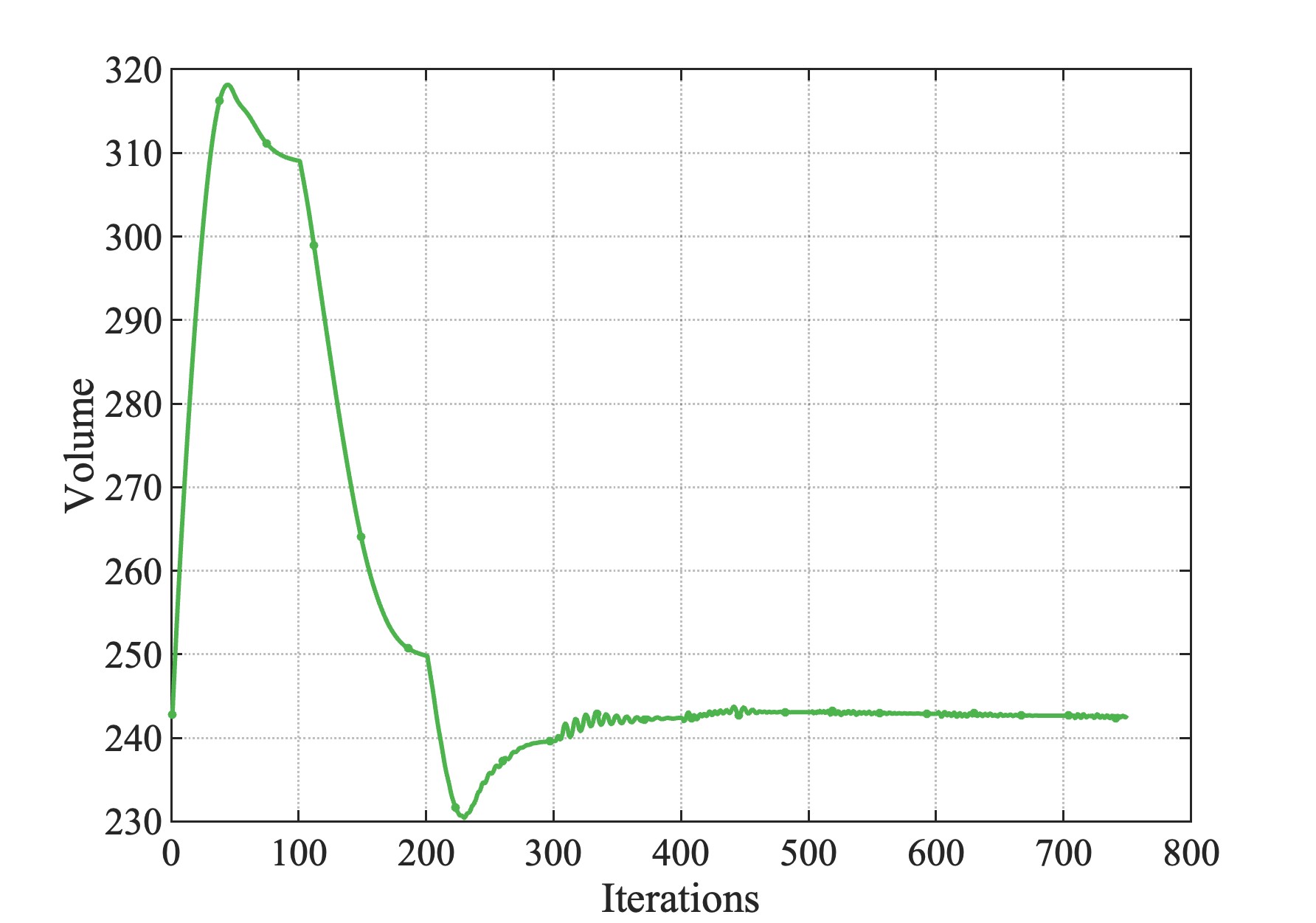}}
\caption{Evolution of the mesh volume over gradient descent iterations during reconstruction by ReMiDi.}
\label{volume_change_over_iterations}
\end{center}
\vskip -0.2in
\end{figure}

Since ReMiDi works by iteratively stretching and compressing the microstructural mesh, updating vertex positions over iterations, one key aspect of its behavior is the change in volume. This is particularly significant because diffusion within a voxel is highly dependent on its volume, determining the extent to which diffusion can occur throughout the microstructure.

The figure~\ref{Latent_space_over_iterations} showcases this volume change over iterations. Around iteration 500, we observe a slowdown in volume change, eventually stabilizing at a fixed value. This trend closely aligns with the convergence observed in spatial loss at a similar iteration, as seen in figure~\ref{spatial_loss_reconstruction}. 

Figures~\ref{signal_loss_reconstruction},~\ref{spatial_loss_reconstruction},~\ref{volume_change_over_iterations} further highlight the inherent ill-posedness of the inverse reconstruction problem. While signal loss converges much earlier, around iteration 250, spatial error and volume converge only around iteration 500.

\subsection{Latent Space Updates during Reconstruction}

\begin{figure}[ht]
\vskip 0.2in
\begin{center}
\centerline{\includegraphics[width=0.5\columnwidth]{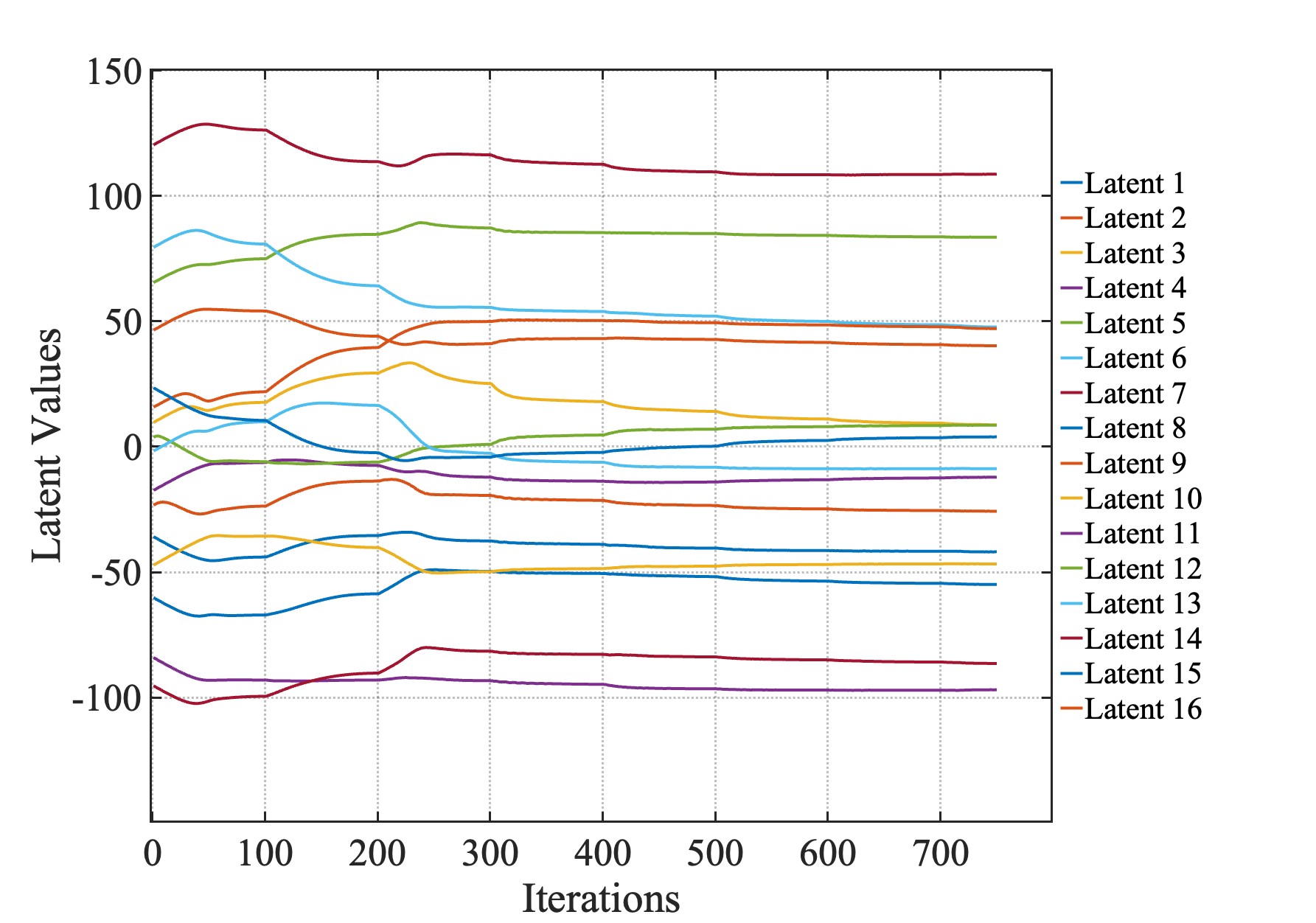}}
\caption{Updating of 16x1 latent space vector representing encoded meshes over iteration by ReMiDi by backpropagation through SAE to latents}
\label{Latent_space_over_iterations}
\end{center}
\vskip -0.2in
\end{figure}

Figure~\ref{Latent_space_over_iterations} illustrates the evolution of the latent space vector (16×1) over iterations. Initially, all 16 values undergo significant changes, but they gradually stabilize around iteration 250. This convergence aligns with the trend observed in figure~\ref{signal_loss_reconstruction}, where signal loss also plateaus at a similar point. Since the $L_2$ loss of the signal is backpropagated to the latent space, the reduction in the error between the reference and computed signals slows down latent space updates over time. To mitigate this early stopping of updates, we use a loss multiplier that scales the loss, ensuring continued iterative updates even when the signal error is minimal.

\end{document}